\documentclass[12pt]{article}
\usepackage{epsfig}
\usepackage{times}
\makeatletter
\DeclareSymbolFont{boldmath}{OML}{cmm}{b}{it}
\DeclareSymbolFontAlphabet{\mathb}{boldmath}
\DeclareMathAlphabet{\Bbb}{U}{msb}{m}{n}

\newcommand{\Tr}{\mathop{\mathrm{Tr}}}

\newcommand{\one}{\mathbf{1}}

\newcommand{\xx}{\mathb{x}}

\newcommand{\uu}{\mathb{u}}

\newcommand{\ww}{\mathb{w}}

\makeatother
\def\be{\begin{equation}}
\def\ee{\end{equation}}

\begin{document}
\title{ High Energy Scattering in the Brane-World and\\
          Black Hole Production}

\author{
I.Ya.Aref'eva\\
{\it Steklov Mathematical Institute, Russian Academy of Sciences}\\
{\it Gubkin St.8, GSP-1, 117966, Moscow, Russia}\\
arefeva@genesis.mi.ras.ru}
\date {$~$}
\maketitle
\begin {abstract}
Black hole production in the  collision of ultra-relativistic
particles in  the   brane-world  approach is considered.
In particular, stability of the brane under collision with
ultra-relativistic particles is discussed.
As a toy model  we consider the
3 dimensional version of the Randall and Sundrum solution
and  show that stability of the brane depends on a choice of
continuation of the solution across the horizon.
In the unstable case  black holes can be produced in the
collision of  a particle with the  brane.
\end{abstract}

\newpage
\section{Introduction}

Main physical questions which are addressed in this letter are the
following
\begin{itemize}
\item
Can two ultra-relativistic particles produce a black hole?
\\
If the answer is "yes", then the following question arises
\item
Is a black hole production observable in theories with low scale
gravity and large extra dimensions?
\end{itemize}

The first question has been   already discussed in the literature
\cite{AGV,Hooft1,VV,AVV}.
In a series of papers Amati, Ciafaloni and Veneziano and 't Hooft
conjectured that black holes occur in the collision of two light particles
at planckian energies. It was argued  \cite{AGV,Hooft1} that at extremely
high energies
interactions due to gravitational waves will dominate all other quantum
field theoretic interactions.
In \cite{AVV}
 the following  scenario for such a process
was proposed. Each of the two
ultra-relativistic particle generates a  gravitational wave
and
 the gravitational waves are  considered as  plane waves.
 Then these plane gravitational waves collide and they
produce a singularity or a black hole.
\be
\label{MBH}
\mbox {{\bf Particles }}~\to ~
\mbox {{\bf  Gravitational Waves
}}~\sim ~
\mbox {{\bf  Plane Gravitational
  Waves}}
~\to ~ \mbox {{\bf Black Holes}}
\ee

To realize this scenario analytically
the Chandrasekhar-Ferrari-Xanthopoulos duality between the Kerr
black hole solution and  colliding plane gravitational waves was used
\cite{AVV,AVVpr}.

A typical parameter for such a process is  the Schwarzschild
 radius $R_S$ of a body  mass m, which is  equal to 
 the energy of colliding particles in
 the center mass frame.
Since   $R_S$ in 4 dimensional case is of order of $M^{-1}_{Pl,4}$ these
process are out
to be observable. However if we accept the brane world scenario
\cite{TeV} where
the fundamental higher dimensional Planck scale  $M^{-1}_{Pl,4+n}
(n>2)$ can be in the TeV range one can expect that such processes
could lead to physical consequences.

For this scenario we need a solution describing
colliding plane gravitational waves in higher
dimensional space-time and we have to find
a black hole geometry in the
collision domain. As a background geometry we consider
brane-world geometry \cite{HW,LOSW}
and  the RS model \cite{RS2} dealing with an infinite extra
dimension.
In this model we live on a brane (domain wall) inside AdS space
and four dimensional gravity is recovered on the brane.
 An analytical description of colliding plane
gravitational waves in $n+4$ dimensional space-time especially in AdS
background is unknown.
We will consider as a toy example the collision of particles
in the 3 dimensional version of the AdS background.
The solution describing   two colliding plane
gravitational waves in $3$ dimensional AdS space-time was found
in \cite{Hooft2,MW}.

We analyze how the presence
of moving particle influences  the brane stability.
We find that a brane in $AdS_3$ due to gravitational
interaction with a particle becomes unstable and it can  split
on  disjoint branes or totally disappears.
This takes place, of course,
only in the case when  a particle can be created.
This case corresponds
to a special continuation of the RS solution across the horizon.
In this case due to the reflection symmetry of the RS solution
the brane in some sense imitates the second particle
and in accordance with  \cite{Hooft2,MW} a black hole can be
created.

One can expect a similar picture in the higher dimensional case.
To support this we use  the one plane wave solution in $AdS_d$
proposed in
\cite{HT,GP,HI,GP2} and argue that to have a black hole production
we have to use a solution that beyond the horizon is pure AdS
with no brane present. If this black hole is created it
is a higher dimensional object.
Phenomenological aspects of such objects have been discussed in
\cite{ADM-R}.
 Black hole formation due to
 gravitational collapse of matter trapped on a brane has been studied
recently in \cite{CHR}.

The paper is organized as following. In Section 2 we remind the
scenario of the black hole creation from \cite{AVV,AVVpr}. In
Section 3 we discuss  changes of geometry of $AdS_3$ in the
presence of moving particles and the influence of these changes on
the brane. In Section 4 some comments about a possible
generalization of 3 dimensional picture  to  higher dimensional
cases are presented.

\section{Colliding Plane Gravitational Waves  and
Black Holes \\ Creation}

Two  main assumptions  of
mechanism
of black holes creation (\ref{MBH}) are \cite{AVV,AVVpr}:
\begin{itemize}
\item
The  transition amplitude for the process
of creation of black hole in the collision of two particles
is determined by the semiclassical
transition amplitude for the process
of creation of black hole in the collision of two
gravitational waves.
\item
Gravitational waves  produced by ultrarelativistic particles
are considered as  plane waves.
\end{itemize}

Saying shortly, the mechanism (\ref{MBH}) means that ultra-relativistic
particles generate  plane gravitational waves
then these plane gravitational waves collide and
produce a singularity or a black hole.
This mechanism uses an idealized picture that  plane gravitational
waves already have been produced by ultrarelativistic particles.
This  idealization based on the fact that ultra-relativistic
particles generate gravitational waves and
any gravitational wave far away from
sources can be considered as a plane wave.
We used this idealized picture because it is difficult to perform
calculations  in the realistic situation.
We assume that plane
waves already have been produced by ultrarelativistic particles
and then consider analytically the process of black hole formation
when the waves collide.

We discuss the process
of creation of black hole in the collision of two plane waves
in  the semiclassical approximation.
Note that black holes  cannot be incorporated into the
theory if we consider quantum field theory in  Minkowski space-time.

There exists a well known class of plane-fronted gravitational waves
with the metric
\begin{equation}
        ds^{2}=2dudv +h(u,X,Y)du^{2}-dX^{2}-dY^{2}
    \label{o}
\end{equation}
where $u$ and $v$ are null coordinates . In particular the
gravitational field of a particle moving with the speed of light
is given by the Aichelburg-Sexl solution \cite{AS}. The metric has
the form
\begin{equation}
ds^{2}=2dudv + E\log(X^2+Y^2)\delta(u)du^{2}-dX^{2}-dY^{2}
        \label{oo}
\end{equation}
and describes a shock wave. It is difficult to find a solution which
describes two sources, except the 3 dimensional case\cite{Hooft2,MW}.
An approximate solution of Einstein
equation  for two particles as the sum of one particle solutions
describes well the scattering amplitude for large impact
parameter, but does not describe non-linear interaction of shock waves
which is dominate in the region of small impact parameter. To analyze
non-linear effects we took, instead of dealing with shock wave,
plane gravitational waves. In some respects
one can consider plane wave as an approximation to  more complicated
gravitational waves, in particular shock waves.
This solution in some sense can be interpreted as an approximation for a
solution of Einstein equation in the presence of two
moving particles.

A particular class of plane waves is defined to be
plane-fronted waves in which the field components are the same at every
point of the wave surface. This condition requires that $h(u,X,Y)$
is a function with a quadratic dependence on $X$ and $Y$.
One can then remove the dependence of h on $X$\ and $Y$ altogether by
a coordinate change.

Classical collision of plane gravitational waves has been the
subject of numerous investigations, see for example
\cite{SKP,CX,FI,Gr}, and it has a remarkably rich structure.
In \cite{AVV} was
used the Chandrasekhar-Ferrari-Xanthopoulos duality between
colliding plane
gravitational waves and the Kerr black hole solution.

\begin{figure}
\begin{center}
\epsfig{file=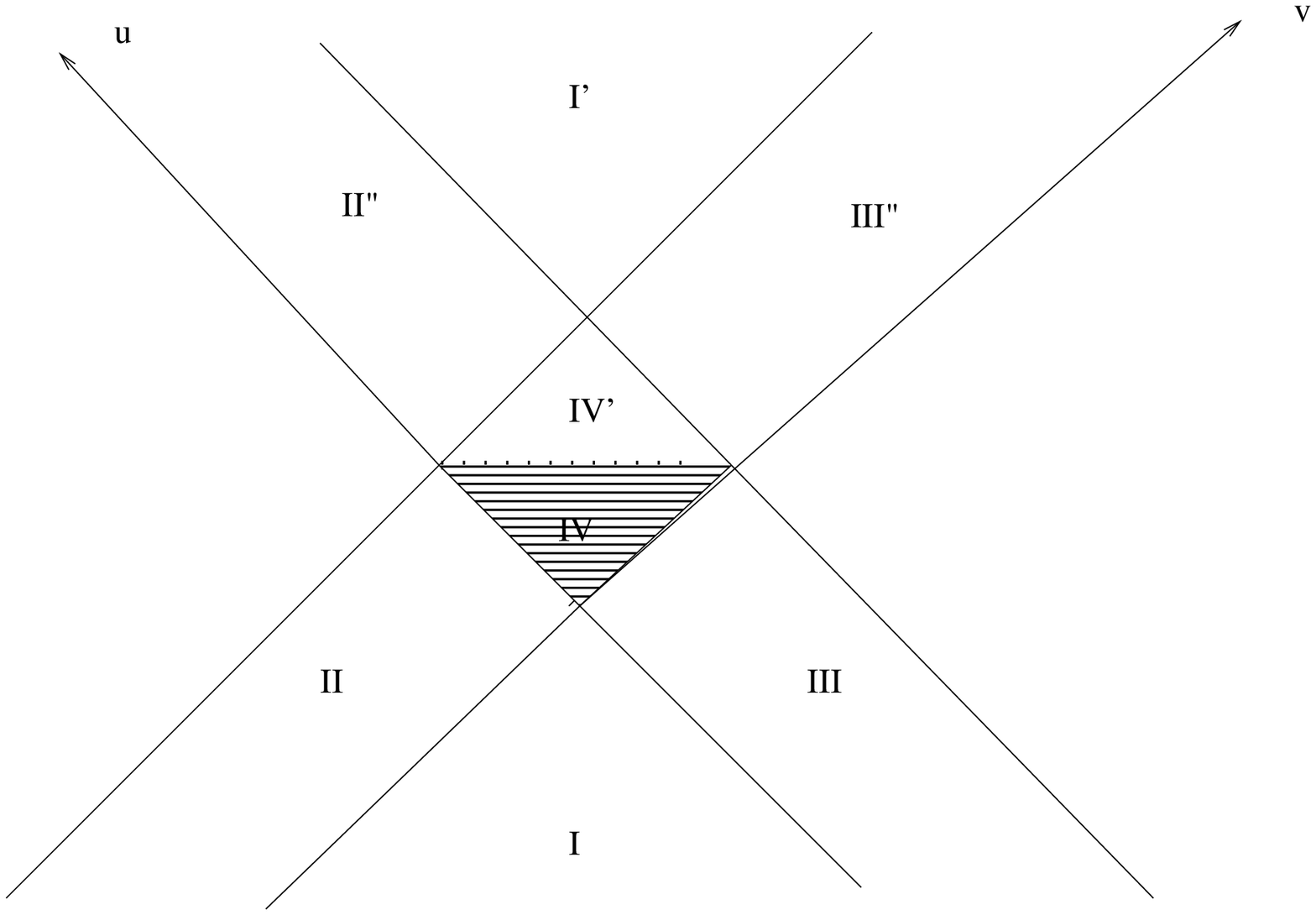,
   width=175pt,
 angle=0
}
$~~~~~~~$
\epsfig{file=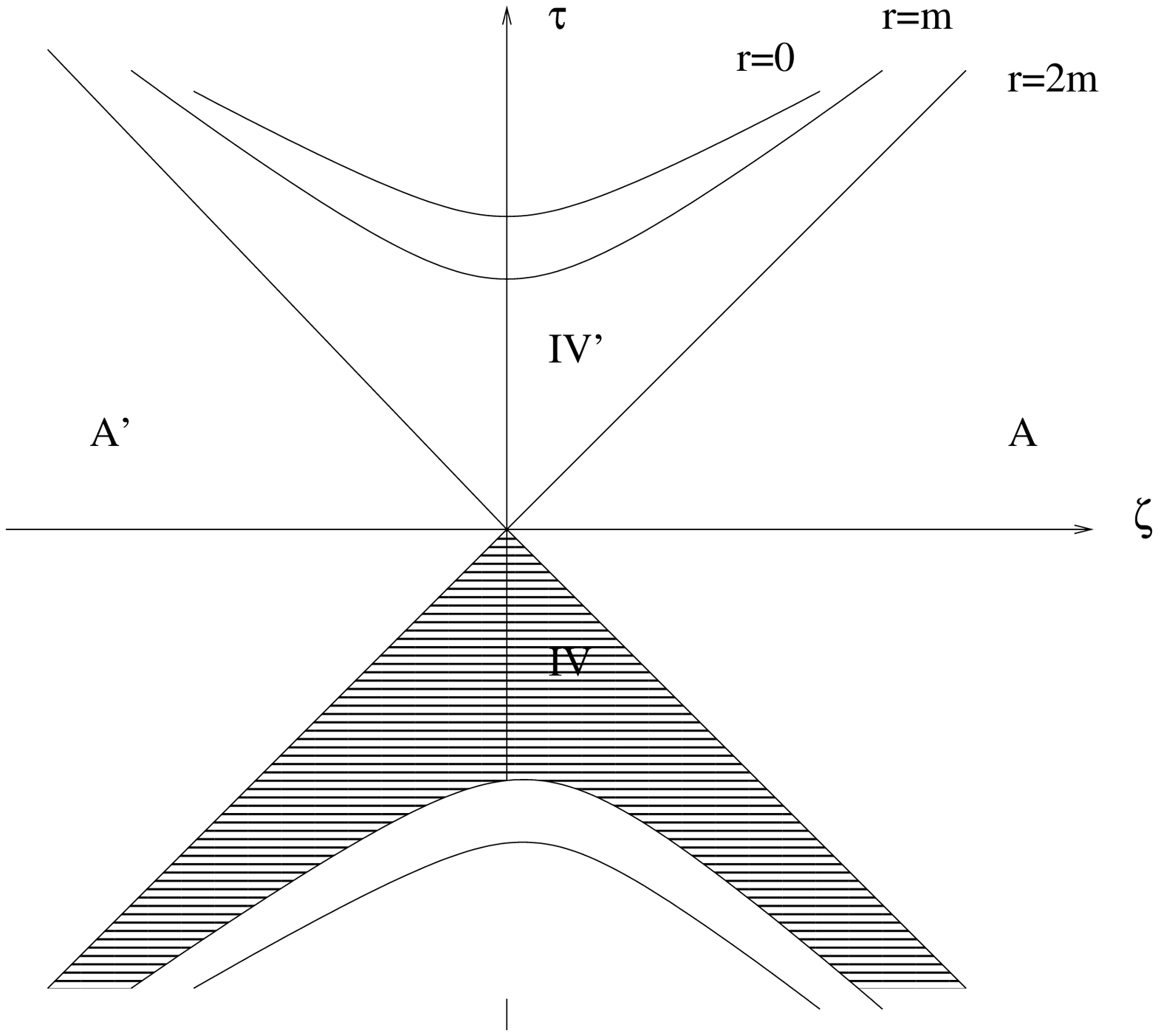,
   width=175pt,
  angle=0
 }
\end{center}
\label{Fig1}
\caption{a) Plane coordinates; $~~$ b) Kruskal coordinates}
\end{figure}

Let us present  a metric corresponding to the collisions of two
plane waves with zero impact parameter. In this case
the colliding plane
gravitational waves produce in the interaction region a space-time that
is isometric to the interior of the Schwarzschild solution.
The metric is given by
$$ds^{2}=  4m^{2}[1+\sin (u\theta (u))+v\theta (v)]  dudv
        $$
\be
\label{6.1}
     -[1-\sin (u\theta (u))+v\theta (v)][1+
        \sin (u\theta (u))+v\theta (v)]^{-1} dx^{2}
 \ee
$$         -[1+\sin (u\theta (u))+v\theta (v)]^{2}
    \cos ^{2}(u\theta (u))-v\theta (v))dy^{2},
$$
where
$u<\pi/2,~~v<\pi/2,~~v+u<\pi/2 .$

  Figure 1 a) illustrates  this solution of the vacuum Einstein equations.
The background
region  I describes a region of space-time before the arrival of
gravitational waves and it is Minkowskian.
Two planes  waves propagate
from  opposite directions along the z-axis.
Regions II and  III contain the approaching plane
waves. In the region IV
the metric (\ref{6.1})
is isomorphic to the Schwarzschild metric.
To see this one can make the following
change of variables from  the plane waves coordinates to the Schwarzschild
coordinates,
$(u,v,x,y) \to (t,r, \theta, \phi)$
defined by,
\be
r=m[1+\sin (u+v)],~~t=x,~~\theta= \pi/2 +u-v, ~~\phi =y/m ,
    \label{6.5}
\ee
or to Kruskal coordinates $\tau, \zeta , \theta, \phi$
\be
\tau =-a(r)\cosh t/4m,~~\zeta =- a(r)\sinh  t/4m,~~
a(r)=(1-r/2m)^{1/2}e^{r/4m}.
    \label{6.6}
\ee
Then one gets
$$
ds^{2}=\frac{32m^{3}}{r}e^{-r/2m}
     (d\tau ^{2} -d\zeta ^{2})
     -r^{2}(d\theta ^{2} +\sin ^{2} \theta d\phi ^{2})
$$
The section of the region  IV bounded
by  $x=0,~ y=0$ corresponds to segment in the Kruskal diagram
and the section of the region  IV
by the plane $x=x_{0},~ y_{0}=0$ corresponds to the hatched region
in the Kruskal diagram figure 2.
The lines corresponding to $r=2m$ (horizon)
apart from the point $(\tau =0,~\zeta =0)$ correspond to the infinite
value of $x$-plane wave coordinate.

The above metric in the $(u,v)$ plane can be extended beyond the event
horizon
$u+v=\pi/2$ in one of two ways.

The first possibility, shown in figure 1 consists in reflecting along the
line $u+v=\pi/2$.
The second one involves in gluing  to the horizon the whole  upper-half
part of the Kruskal diagram.

Both extensions are solutions of Einstein equations.
There is a-priori non-zero probability
to get a finite state corresponding to a black hole or two
outgoing plane waves. Calculations of the
probabilities for these processes
in the semi-classical approximation are performed in \cite{AVV}.


\section{Brane-particle collision in $AdS_3$}
In this section we consider the solution of Einstein equations
describing interaction of brane and particle in the $AdS_3$ spacetime.

To describe a moving particle in $AdS_3$ it is convenient to use the
global coordinate system $(\tau,r,\phi)$
and the matrix representation
\be
\label{mat}
\xx=x_{-1}{\one}+x^{a}\gamma_a,~~\det \xx=1,
\ee
where  $r$ is a radial coordinate $~0< r\leq 1$, $~0\leq \varphi < 2\pi$
 is an angular coordinate, $\tau$  is a real time
coordinate and
\be
\label{gamma}
\gamma_0 = \pmatrix{ 0 & 1 \cr -1 & 0 } , \qquad
 \gamma_1 = \pmatrix{ 0 & 1 \cr 1 & 0} , \qquad
 \gamma_2 = \pmatrix{ 1 & 0 \cr 0 & -1 }.
\ee
The metric is

\be
\label{rho}
 d s^2 =  \frac{1}{2}\Tr ({\bf x}^{-1}d{\bf x}~{\bf x}^{-1}d{\bf x})=
 \Big(\frac{2}{1-r^2}\Big)^2 \,
    \big( d r^2 + r^2 ~d \varphi^2 \big)
          - \Big(\frac{1+r^2}{1-r^2}\Big)^2 ~ d \tau^2.
\ee
$AdS_3$ space can be  considered as a Poincar\'e
disc evolving in time.
To construct a space-time  containing
a point particle according to \cite{Hooft2,MW}
 one has to fix the holonomy, say
 \be
 \label{hol}
 \uu = \one + \tan\epsilon \, (\gamma_0 + \gamma_1), \qquad
     0 < \epsilon < \pi/2,
 \ee
and find a curve $\ww _-$ in the $\tau$-plane
such that  its image $\ww _+$ under a spatial isometry,
$\uu \ww _+\uu ^{-1}=\ww _-$ lies in the same $\tau$-plane.
 Then one has to cut out the wedge between these lines and
to identify the faces according to the isometry.
A world line of the particle
is the set of fixed points of the isometry.
The resulting space-time manifold has a constant curvature
everywhere except
on the world line. The curves $\ww _{\pm}$ are given by
\be
\label{ww}
\frac{2r}{1+r^2}\sin (\epsilon \pm \varphi)=\sin \tau\sin \epsilon
\ee

\begin{figure}[h]
\begin{center}
\epsfig{file=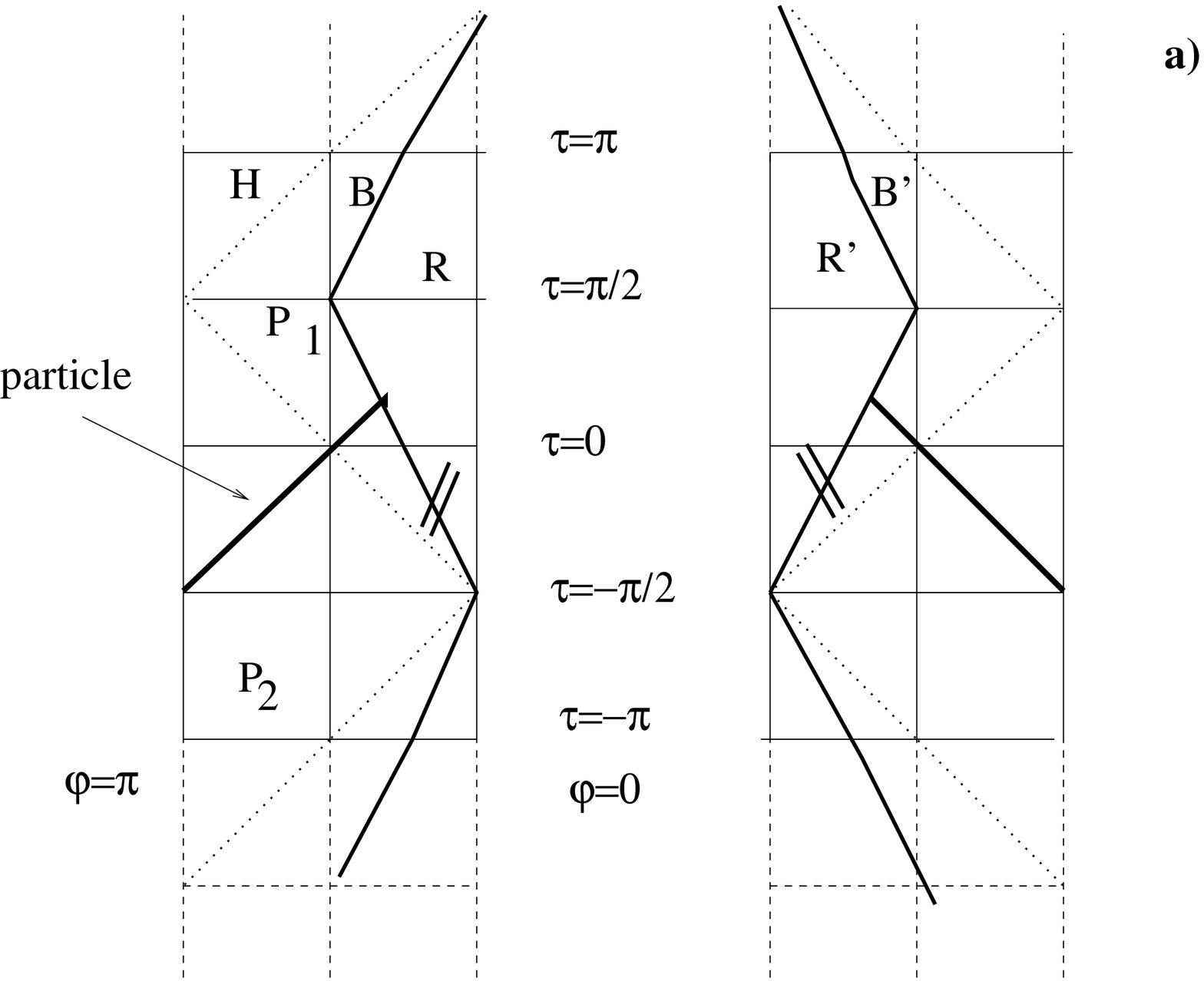,
  width=200pt,
  angle=0
 }
\epsfig{file=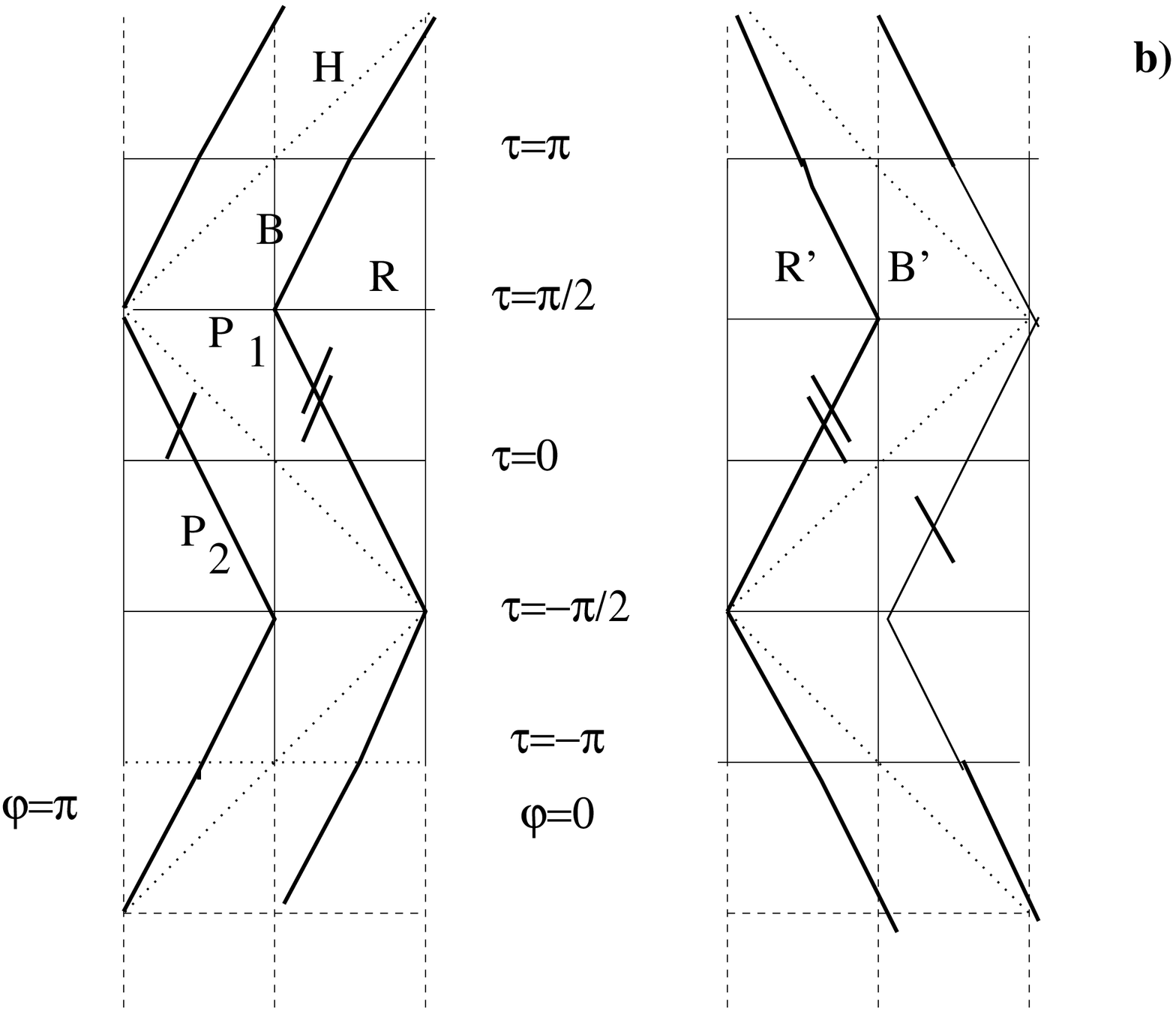,
  width=200pt,
  angle=0
 }
\end{center}
\caption{Penrose diagrams for $AdS_3$ with a brane.
A brane $B$ is located in the region $P_1$.
Diagrams a) and b) show the different continuations across
the horizon H.
a) there are no branes in the region $P_2$.
b) there is a brane B in the region $P_2$.
Identifications are shown by $"/"$ and $"//"$.
$R$ and $R'$ denote the removed regions.}
\label{pen2}
\end{figure}

\begin{figure}[h]
\begin{center}
\epsfig{file=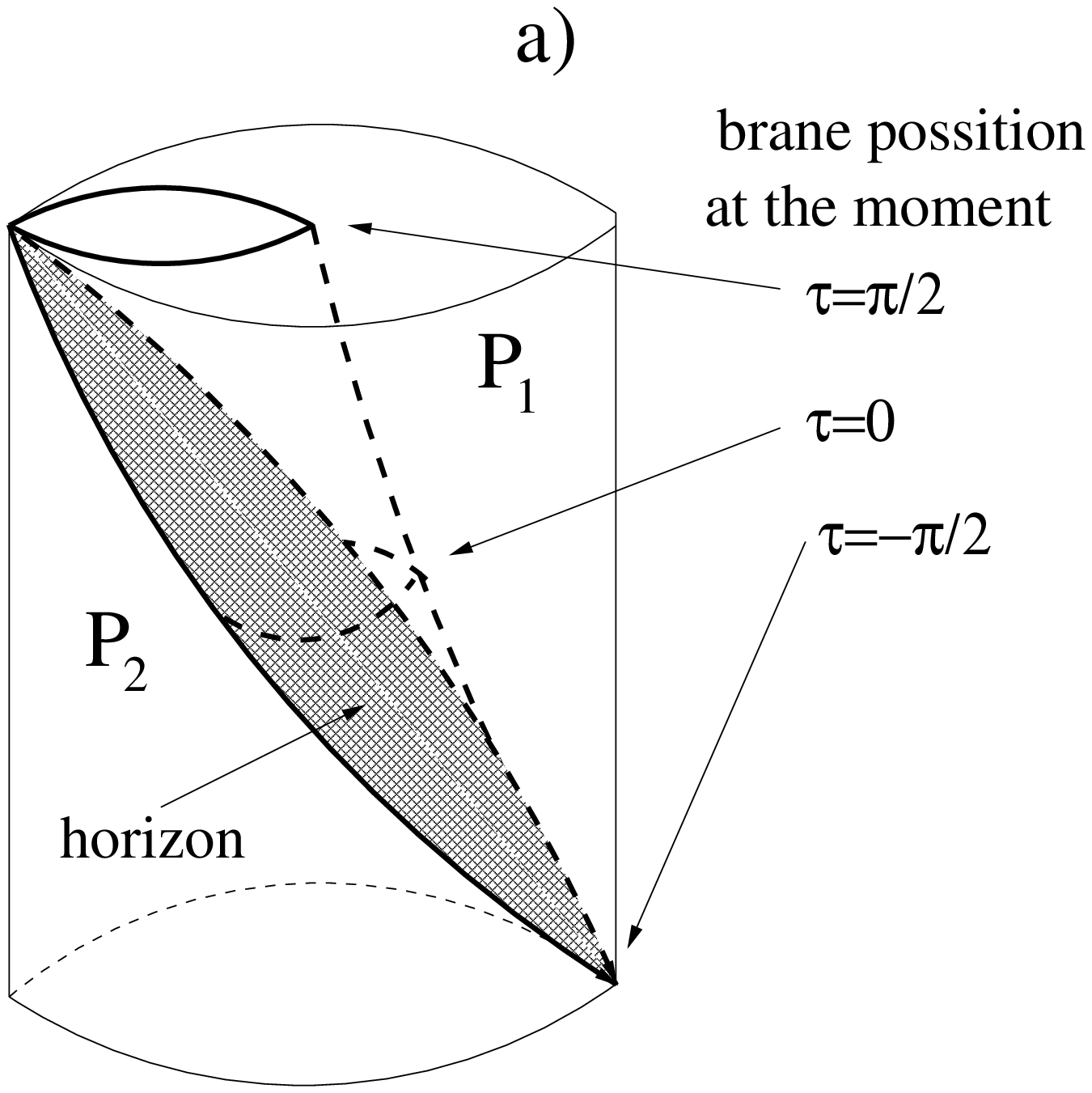,
  width=150pt,
  angle=0
 }$~~~~~~~~~~~~~$
\epsfig{file=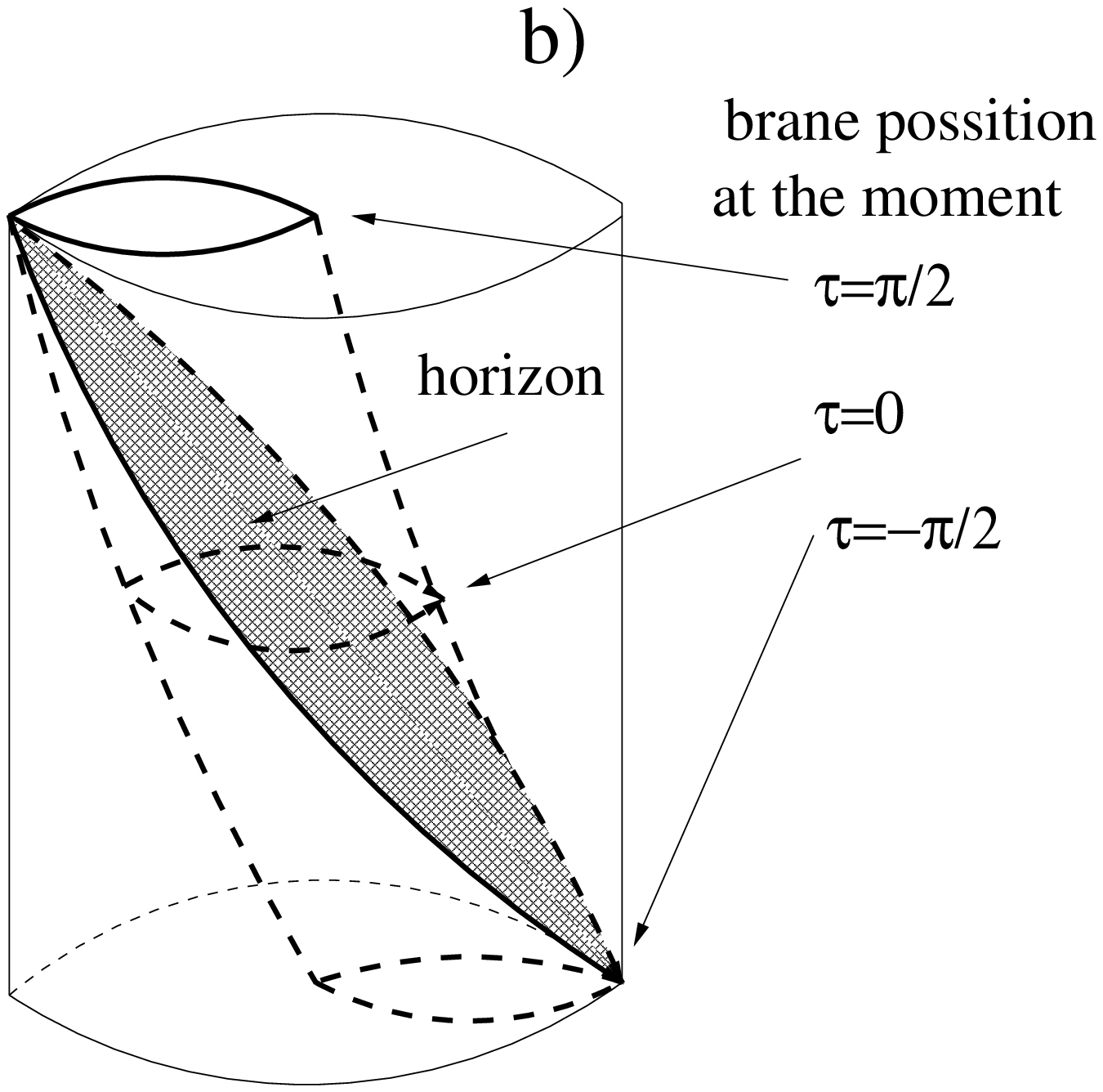,
  width=150pt,
  angle=0
 }\end{center}
\caption{Three dimensional picture of the brane. $AdS_3$ is
 displayed as the
interior of the cylinder. The horizon $H$ (the shared cut) divides $AdS_3$
into two regions $P_1$ and $P_2$, each of which is covered by a set of
Poincare coordinates. a) and b) show the different continuations across the
horizon}
\label{3d}
\end{figure}

Let us now consider the tree dimensional version of the Randall
and Sundrum (RS) model \cite{RS2} which deals with a brane
 located
at   $y=1$,
where y is one of the set
of the Poincare coordinates describing $AdS_3$.
RS slice AdS along the surface  $y=1$, remove the portion
$0<y<1$ and assume $Z_2$ reflection symmetry at
the boundary surface. In the global coordinate system
$(t,r,\phi)$ the 2 dimensional surface $y=1$
is described by the equation
\be
\label{brane}
\cos \phi =\frac{1-r^2-(1+r^2)\sin \tau}{2r}
\ee

Let us note that there are several ways to analytically continue the
RS solution across the horizon (see \cite{CSrep} and references therein).
Two obvious choices of continuation are shown in figure \ref{pen2}.
For the first continuation (figure \ref{pen2}a) there is no brane beyond the
horizon.
For the second one (figure \ref{pen2}b) there is a brane beyond the
horizon.

In figure  4 the brane positions under the assumption of the 
global structure
presented  in figures \ref{pen2}a) and \ref{3d}a)
are shown for different values of $\tau$.
We consider here the brane positions for the time between $\tau=-\pi/2$
and $\tau=\pi/2$. We see that the brane at the initial moment is
just a
point on the Poincare disk and becomes a circle of 1/2 radius in the last
moment.

One can consider the brane located at $y=a$
 \be
\label{brane_a}
\cos \phi =\frac{(1-r^2)a^{-1}-(1+r^2)\sin \tau}{2r}
\ee
In the case of $a\neq 1$
the brane positions
between $\tau=-\pi/2$
and $\tau=\pi/2$  are shown in figures 4b and 4c.
For all values of $a$ at the initial moment the brane  is
just a
point on the Poincare disk. For $a$ small enough
it becomes a closed curve located near the boundary of the Poincare disk
  in the last
moment. For large  $a$  and $\tau=\pi/2$ the brane is a curve located near
$\varphi =\pi$

\begin{figure}[t]
\begin{center}
\epsfig{file=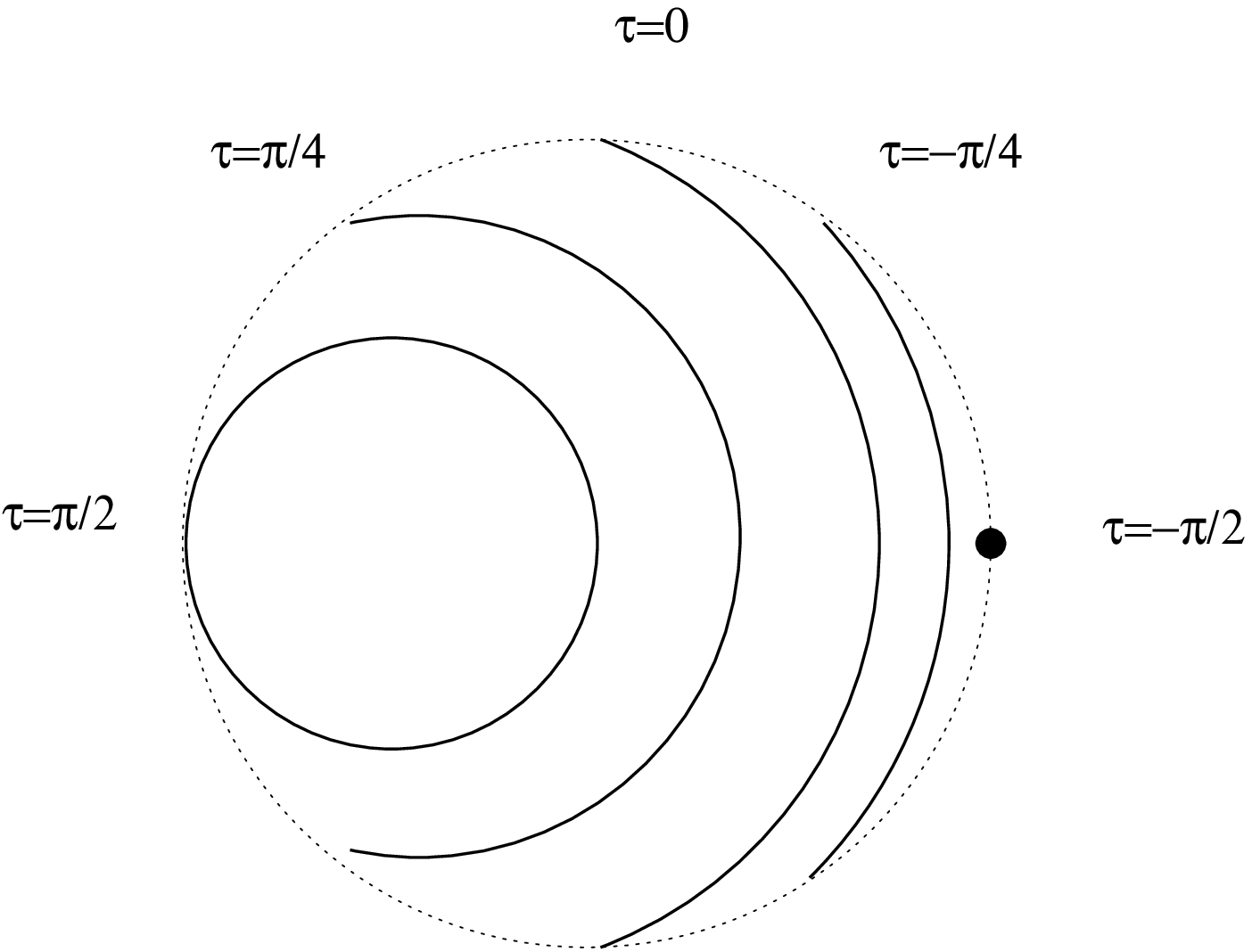,
   width=100pt,
  angle=0}$~~~$
\epsfig{file=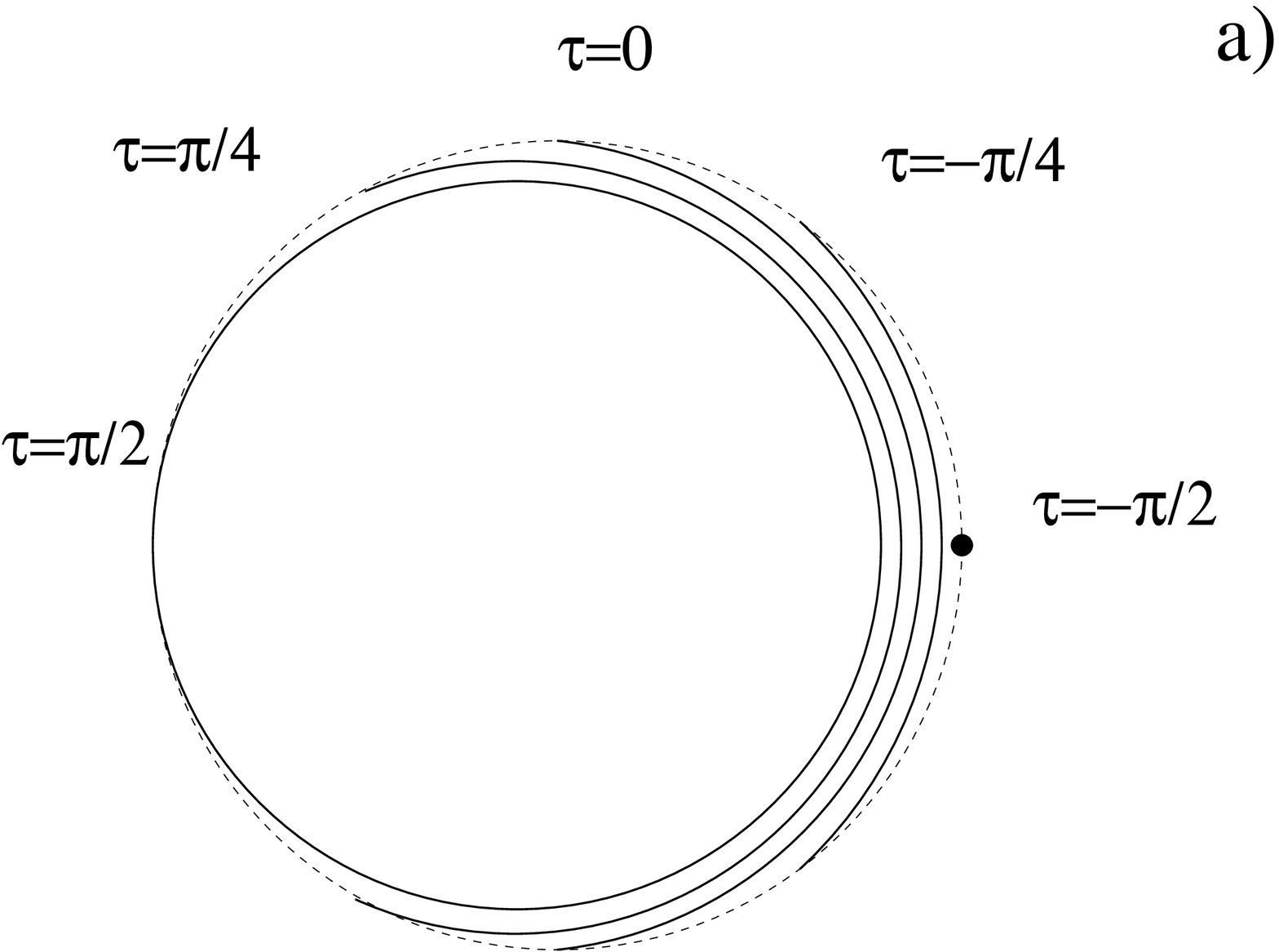,
   width=100pt,
  angle=0
 }
 $~~~$
 \epsfig{file=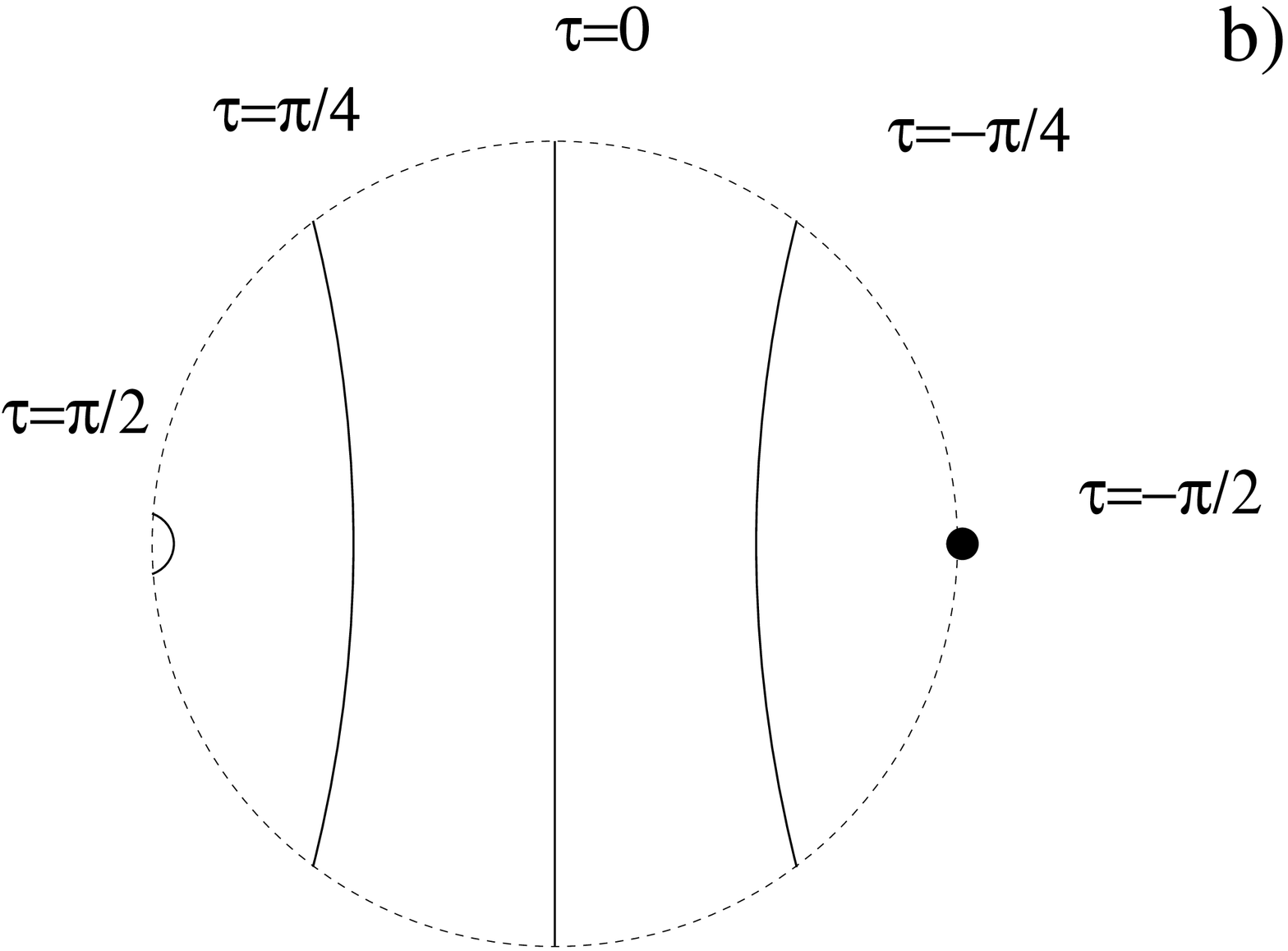,
   width=100pt,
  angle=0
 }
\end{center}
\label{Fig4}
\caption{Brane positions on the Poincare disk in  times $-\pi/2\leq
\tau \leq \pi/2$: a) $a=1$,$~~$ b)$a<<1$;$~~$ c) $a>>1$}
\end{figure}

\begin{figure}[h]
\begin{center}
\epsfig{file=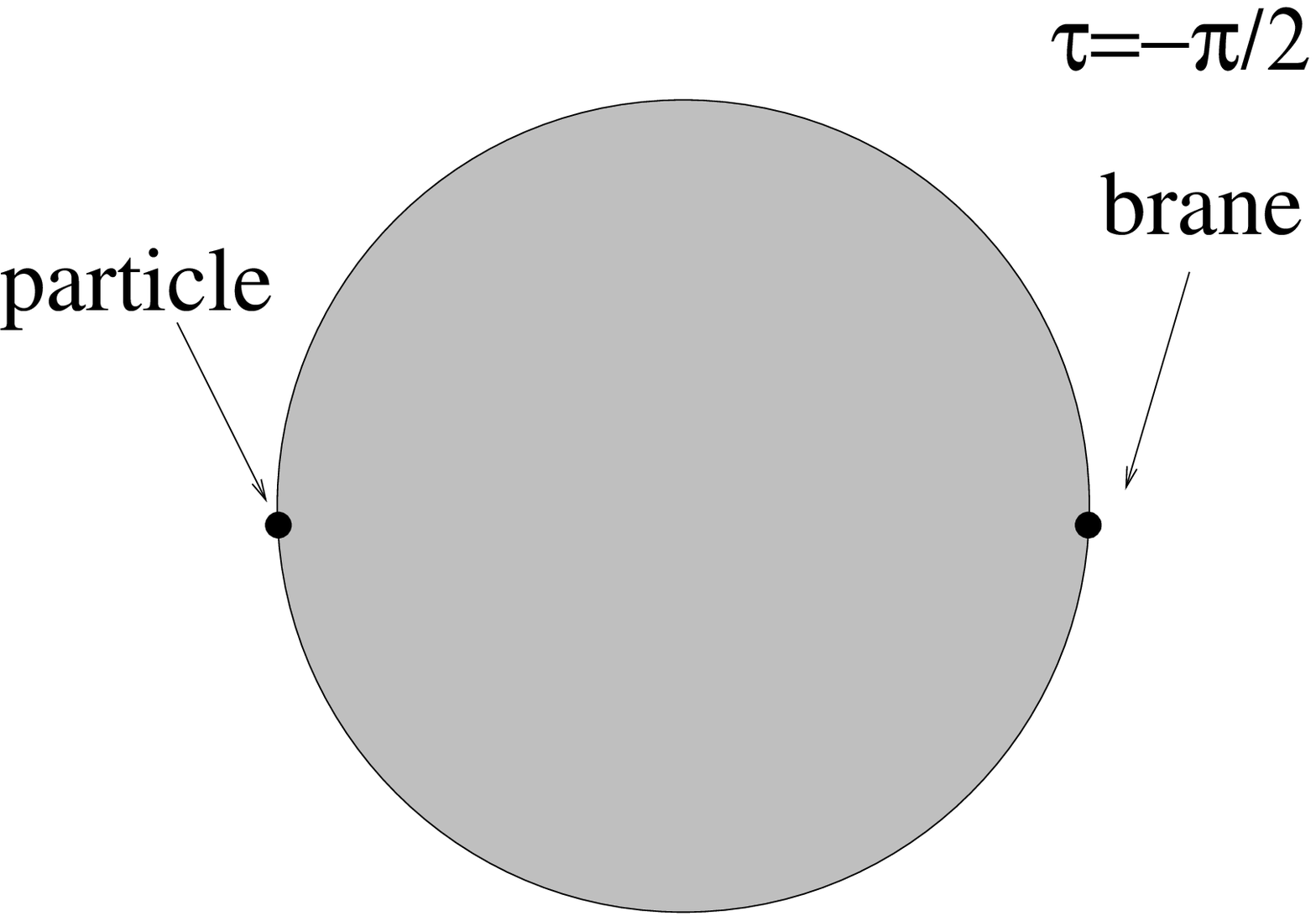,
           width=120pt,
           angle=0}$~~~~~~~~~~~~~$
\epsfig{file=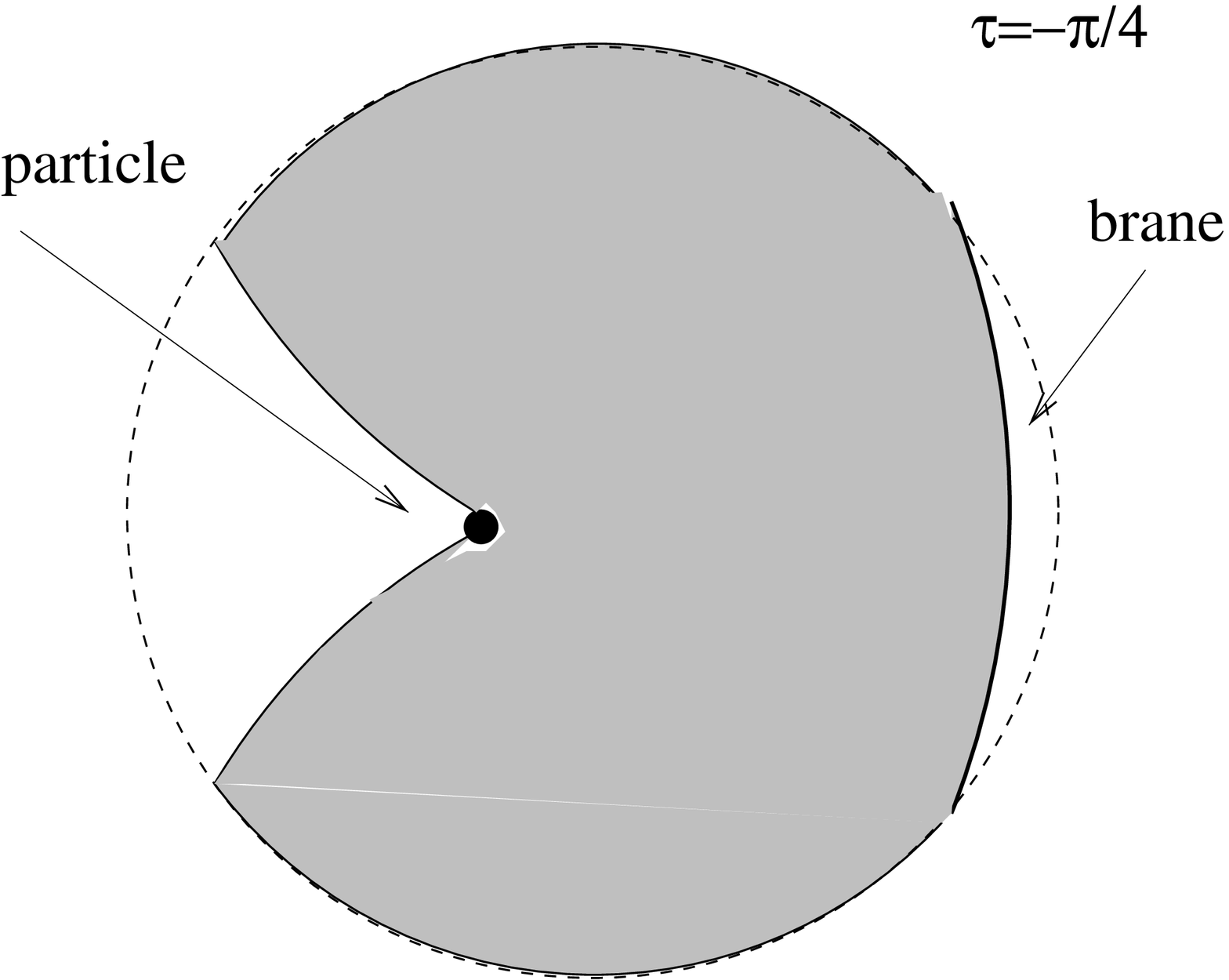,
         width=100pt,
                      angle=0}$~~~~~~~~~~~~~$
\epsfig{file=
         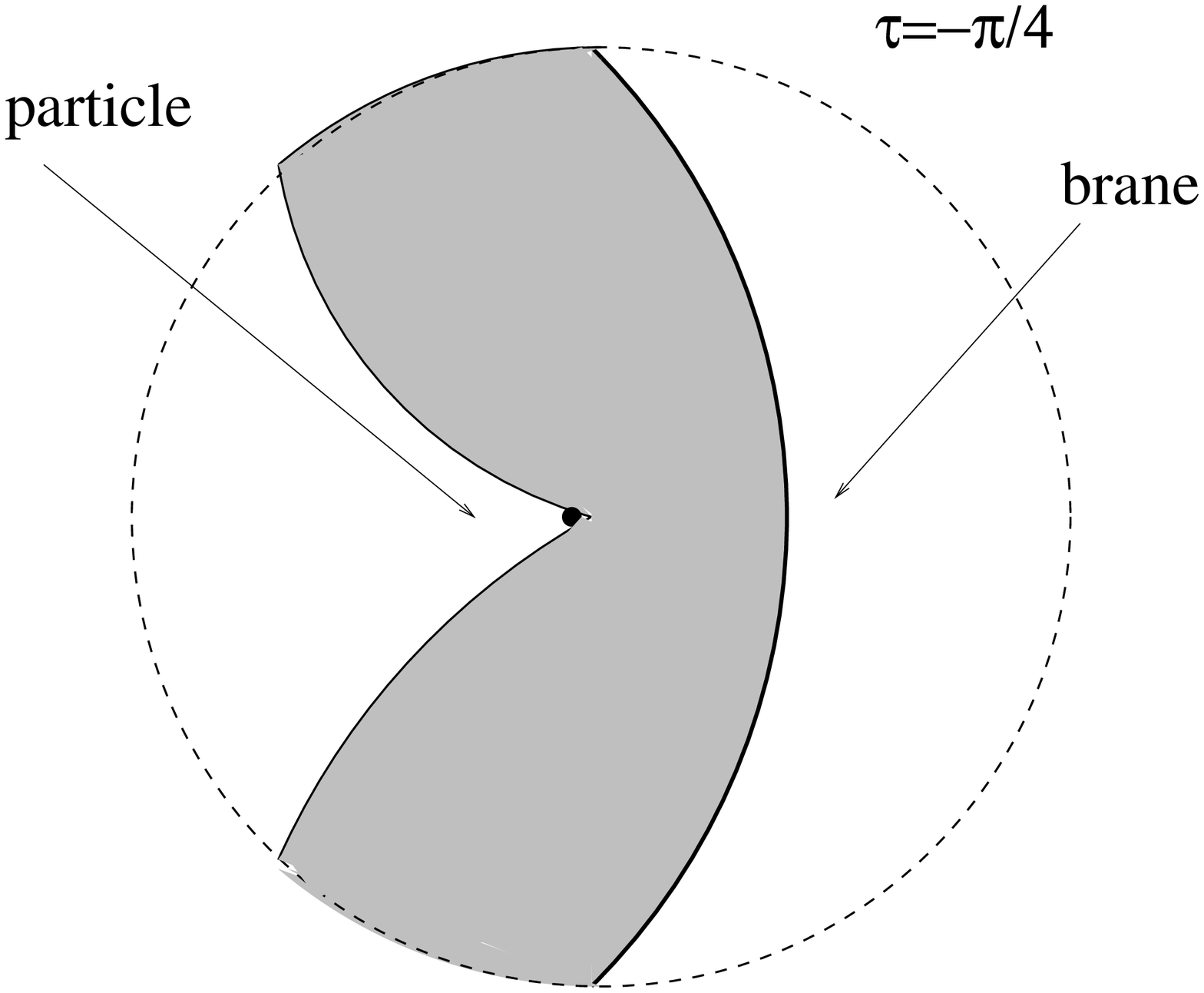,
         width=100pt,
                                 angle=0}
\end{center}
$$~$$
\begin{center}\epsfig{file=
         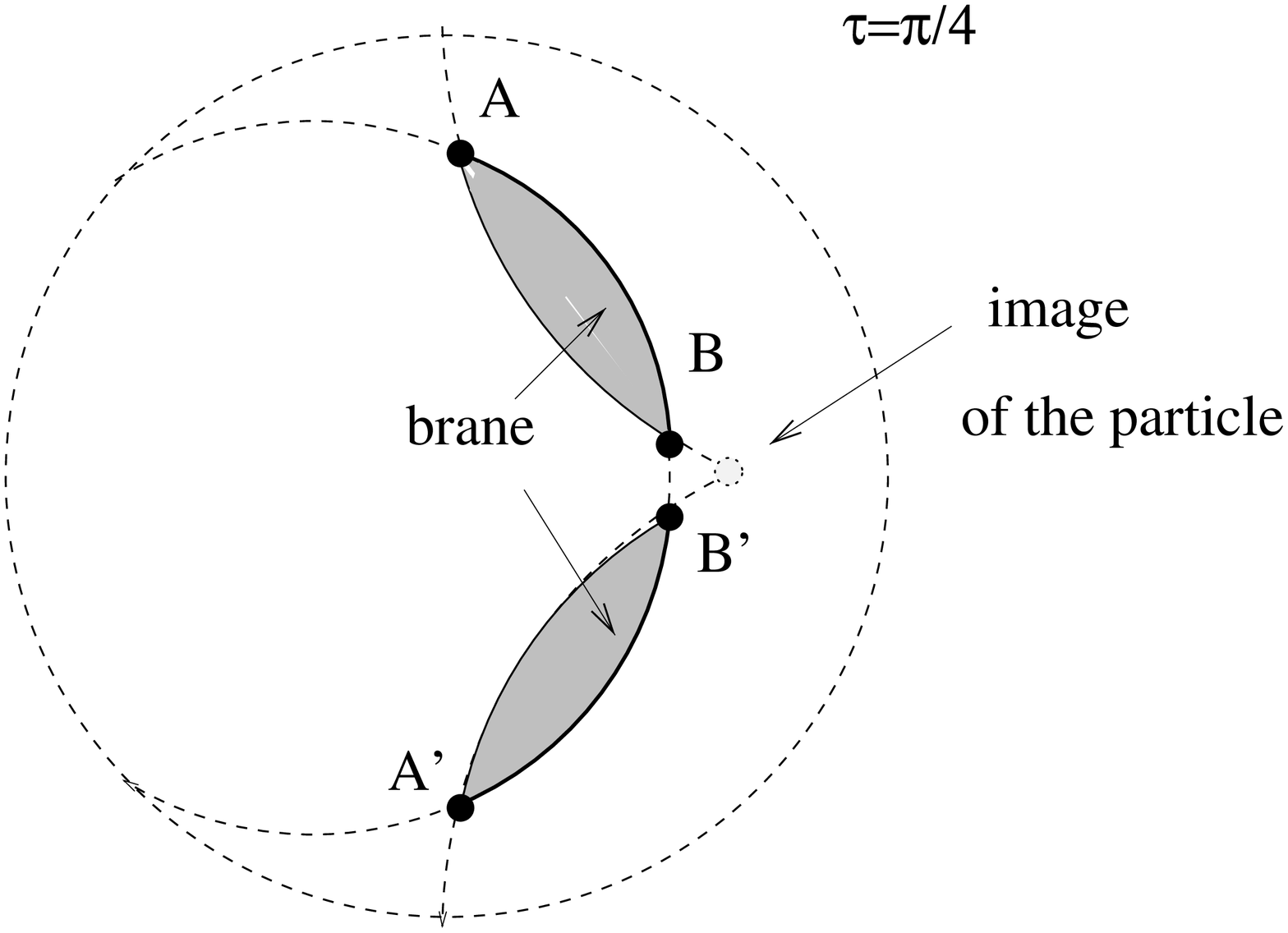,
         width=100pt,
                          angle=0}$~~~~~~~~~~~~~$
\epsfig{file=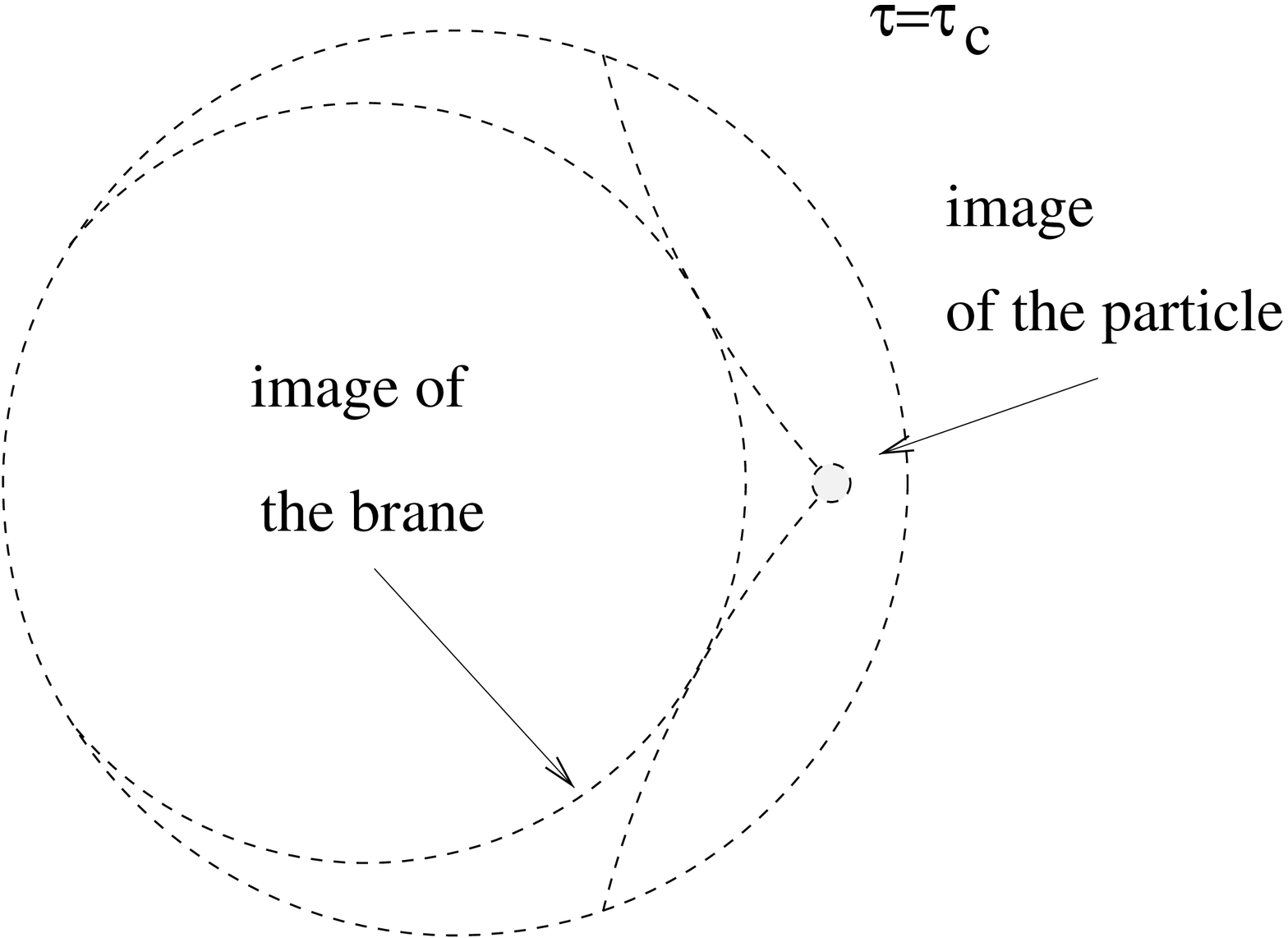,
   width=100pt,
  angle=0
 }$~~~~~~~~~~~~~$
 \epsfig{file=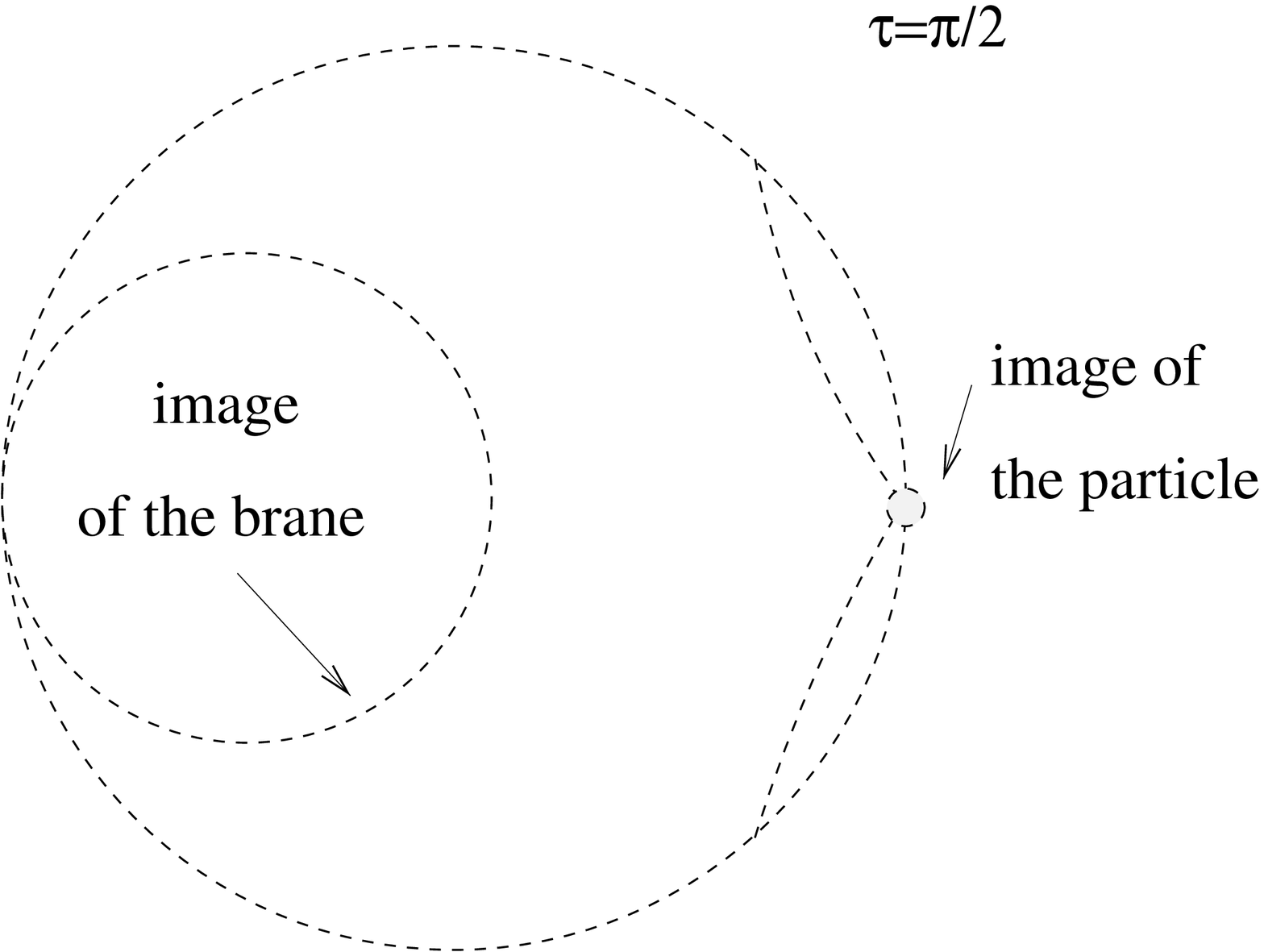,
   width=100pt,
  angle=0
 }
\end{center}
\label{Fig5}
\caption{Brane positions on the Poincare disk with wedge cuts ($\epsilon=
\pi/4)$ in  times
$-\pi/2\leq
\tau \leq \pi/2$}
\end{figure}

Let us consider the collision of the brane and the particle. We
assume that at the initial moment $\tau=-\pi/2$ the particle is at
the point $\varphi =\pi, ~r=1$ (the case of the symmetric initial
position of the particle and the brane).  
The pictures of positions of the brane and the wedge
in subsequent moments of time  $-\pi/2\geq \tau\geq -\pi/2$ are
presented on figures 5 and 6. In dependence of
holonomy (\ref{hol}) characterizing the moving particle,
there are two different cases. 
 Both of
them schematically are presented on figures 5 and 6. In the first
case the "removing" part of the Poincare disk at the moment $\tau
=\tau _c$ is large enough to place there the domain with $y\geq
1$.
Note that since the boundaries
of wedge are unidentified (points A and A' as well as B and B' are
identified) the brane  on figure 5 is in fact connected.
 At the moment $\tau=\tau_c$ the brane  is totally in a
region that we have to cut out and the brane disappears.
To have this picture we have to take $\epsilon$  near to $\pi/2$.
In the second case the brane  does not disappear between $\tau=-\pi/2$
and $\tau=\pi/2$.
This picture takes place for  $\epsilon $ near to $0$.

A case of a nonsymmetric
position of the particle and the brane in the initial
moment $\tau=-\pi/2$
is presented on figure 7. Note that here at the moment
$t=\pi/4$ the brane is composed from two
pieces  $CA$ and $B'D$ (point $A $ is  identified with $A'$
but not with  $B'$).
At the moment $\tau =\tau _0$ the word line of the particle crosses
the brane and the brane splits on tree pieces. Later one of these
pieces disappears.  Two pieces  of the brane  are pasted
to one brane at the moment  $\tau =\pi/2$.

The above consideration demonstrates
that the brane configuration is unstable under possible collisions with
particles. Note that this consideration
assumes the global structure \ref{pen2}a.

Let us now consider the second way of continuation across the  horizon.
The global structure of space time is shown on \ref{pen2}b.
In  this case one deals with  further branes beyond the horizon
and there is no timelike infinity in the space time.
Hence there is no "place" from which a particle can start its evolution.
This shows a stability of the RS solution under assumption of many
branes.

\section{Discussion and Conclusion}
In the  4 dimensional case  the process  of the black hole creation
(\ref{MBH})
takes place for the impact parameter $b$ smaller than
 the Schwarzschild radius of the black hole with the  mass equals to
the energy of colliding particles in the center mass  frame.
Analogously we expect
that the same process dominates in the $n+4$ dimensional case for
\be
\label{bn}
b<R_{S,n+4},
\ee
where $R_{S,n+4}$ is the Schwarzschild radius of the  $4+n$
dimensional black hole with mass $m$.
The mass $ m$ is equal to
the energy of colliding particle in the center mass  frame.
The Schwarzschild radius of the $4+n$
dimensional black hole of mass $m$ is given by \cite{RM}
\be
\label{Rn}
R_{S,n+4}=c_{n+4}~(\kappa _{4+n}m)^{1/n+1}
\ee
there $c_{4+n}==(\frac{8\Gamma ((n+3)/2)}{(n+2)\pi ^{n+1/2}})^{1/n+1}$.
Using  that
$\kappa _{4+n}\sim M^{n+2}_{Pl,4+n}$
we have the bound
\be
\label{bna}
b<c_nM_{Pl,n+4}^{-1}~(\frac{E}{M_{Pl,n+4}})^{1/n+1},
\ee
where $E$ is the energy of colliding particles in the center mass  frame.
Therefore, if one adapts the scenario of \cite{TeV}
and finds an analog of the metric (\ref{6.1}) then   one can conclude
that the black hole production  takes place at the Tev scale.
This means that one has a very strong restriction according to which
 processes with transverse
momenta larger than $R_{S,n+4}^{-1}$
should be completely absent.
We see an interesting feature of the bound (\ref{bna}).
Since we expect that $n$ is large enough, say $n=6$,
the right hand side  of this inequality does not depend very much on the
energy of colliding particles.

Let us note once again that to realize
(\ref{MBH}) in higher dimensional case one has to find
a solution  describing collision of
gravitational waves.
This is still an open problem for
$n>0$. Moreover,
within the framework of low scale  quantum gravity scenario
one has to solve a problem of colliding waves in a particular compactified
space.
However, by analogy with the 3 dimensional case one can expect
that the role of the second plane wave can be played by the brane within
the RS scenario. The one plane wave in the $AdS_d$ background was
found for $d=4$ in \cite{HT,GP,GP2} and for $d>4$ in \cite{HI}. The metric
has a simple form in plane wave coordinates
\be
\label{ads-pl}
ds^2=\frac{dUdV-\eta _{\alpha \beta}(dZ^{\alpha}+
U\theta (U)H_{\alpha\gamma}dZ^{\gamma})(dZ^{\beta}+
U\theta (U)H_{\beta\gamma'}dZ^{\gamma'})}
{[1-(UV-\eta _{\alpha \beta}Z^{\alpha}Z^{\beta}+U\theta (U)G]^2},~\alpha,
\beta=1,..d-2.
\ee
This metric for negative values of $U$
reproduces the pure $AdS_d$ metric.
The plane wave coordinate U is related with the global coordinates
via
\be
U=\frac{\tan \rho ~n_1-\sec\rho \sin \tau }{1+\sec \rho \cos \tau}
\ee
where $n_1$ is the first component of $d-2$ dimensional unit vector.
This metric describes a metric in the presence of a massless particle which
moves along the null geodesic $U=0$ in $AdS_d$
background. In the Penrose diagram (see figure \ref{pen2}a)
this looks like as the world line of the particle in the 3 dimensional
case. This consideration supports an analogy with the 3 dimensional
case, although to be sure a more detailed analysis is needed.

To summarize, in this letter we discussed the application
of the mechanism of the black hole production
from colliding plane gravitational waves in 4 dimensional space time
to the Tev energy scattering
in the n+4 dimensional spacetime in the presence of a brane.
We have shown that the brane could be unstable in the presence of
gravitational waves and the black hole can be formed.
It was shown that the brane is stable within the many branes version of
the RS solution in the three dimensional case.
We also noted that a bound on transverse
momenta of completely absent processes  does not essentially
depend on the energy of colliding particles.

\section*{Acknowledgments}

I would like to thank I.V.Volovich
for discussions.  This work was supported in part by
INTAS grant 96-0698,   by RFFI grant99-01-00166  and by grant
for the leading
scientific schools 96-15-96208.

\newpage

\begin{figure}[t]
\begin{center}
\epsfig{file=bp1.eps,
           width=120pt,
           angle=0}$~~~~~~~~~~~~~$
\epsfig{file=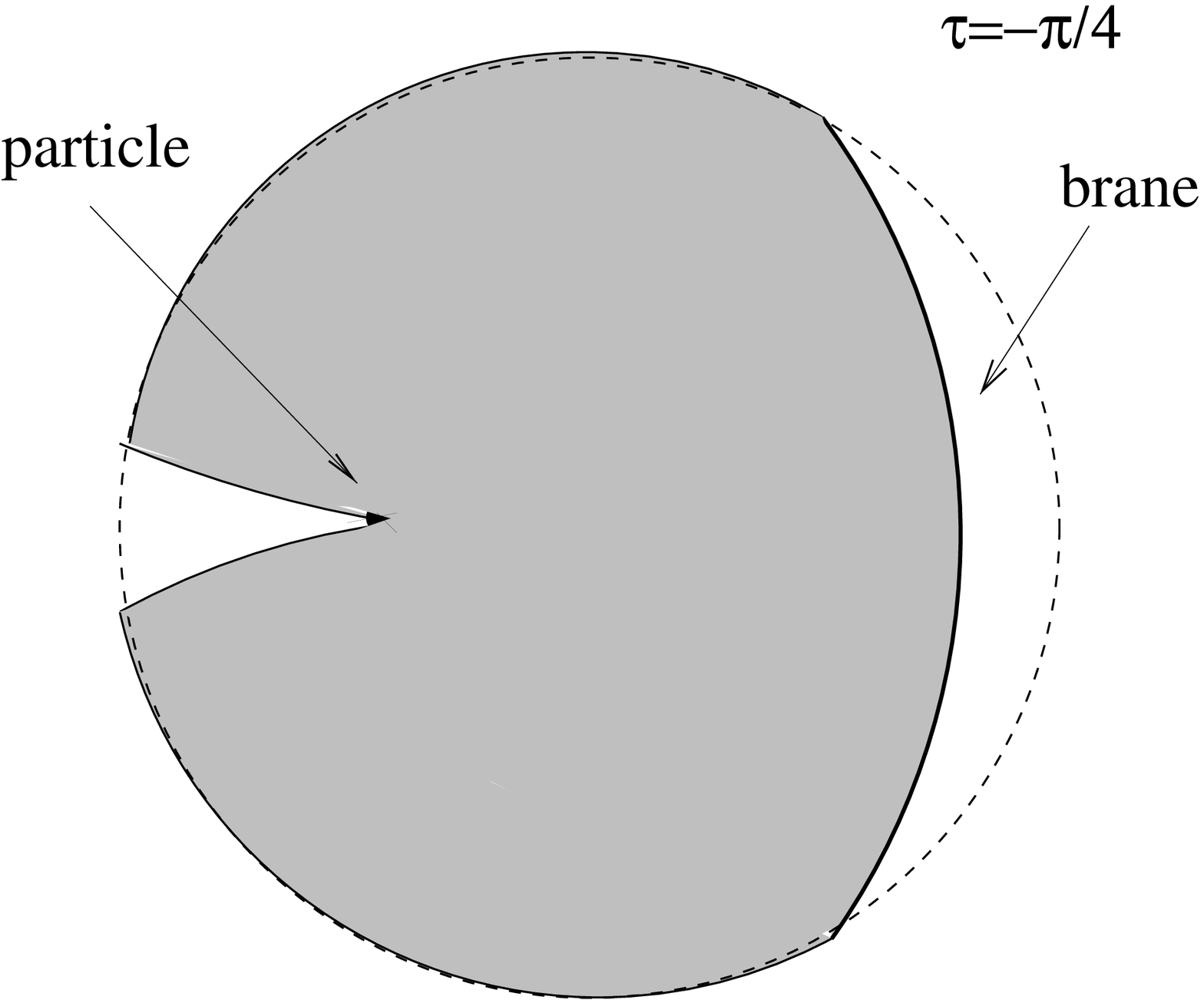,
         width=100pt,
                      angle=0}$~~~~~~~~~~~~~$
\epsfig{file=
         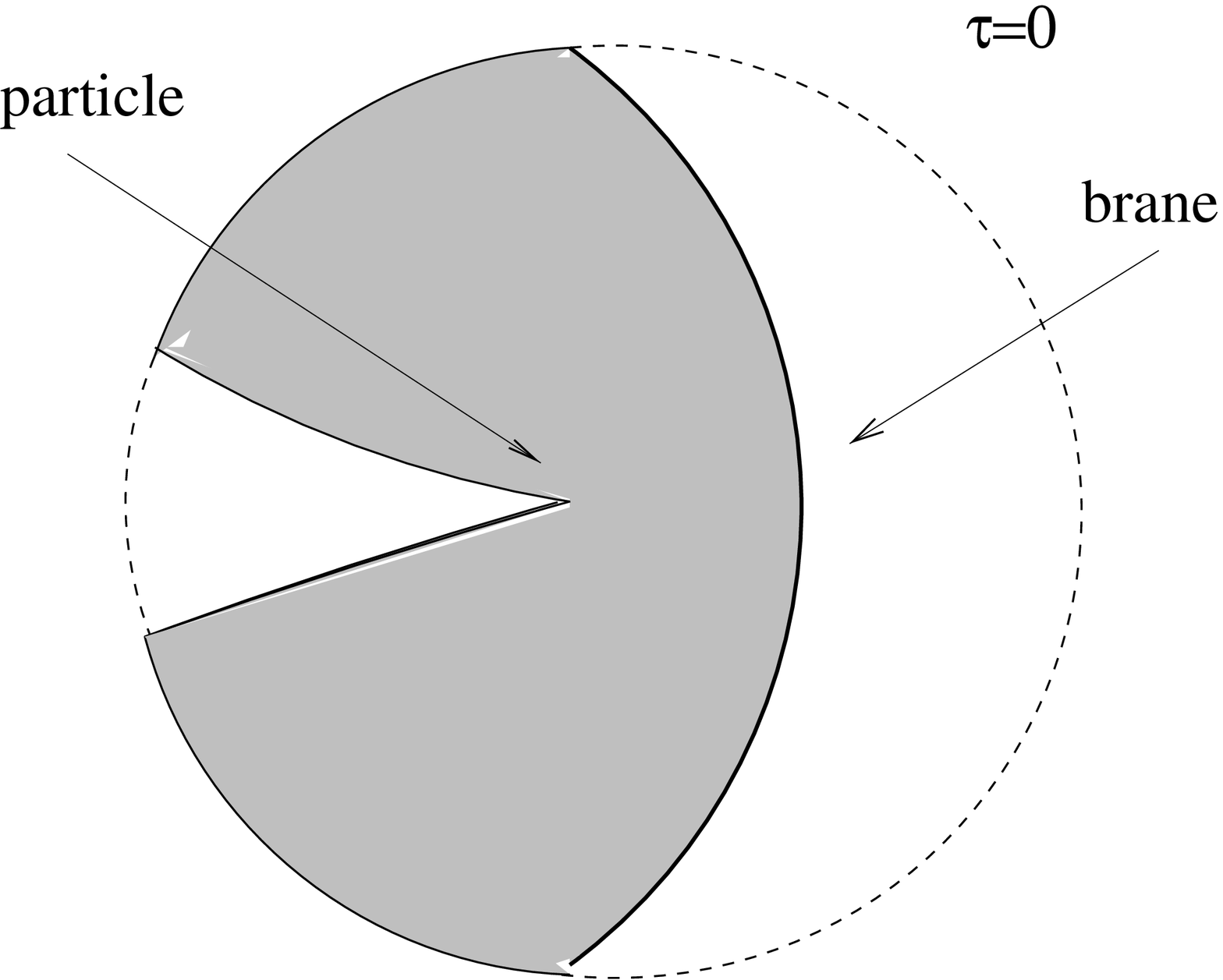,
         width=100pt,
                                 angle=0}
\end{center}
$$~$$
\begin{center}\epsfig{file=
         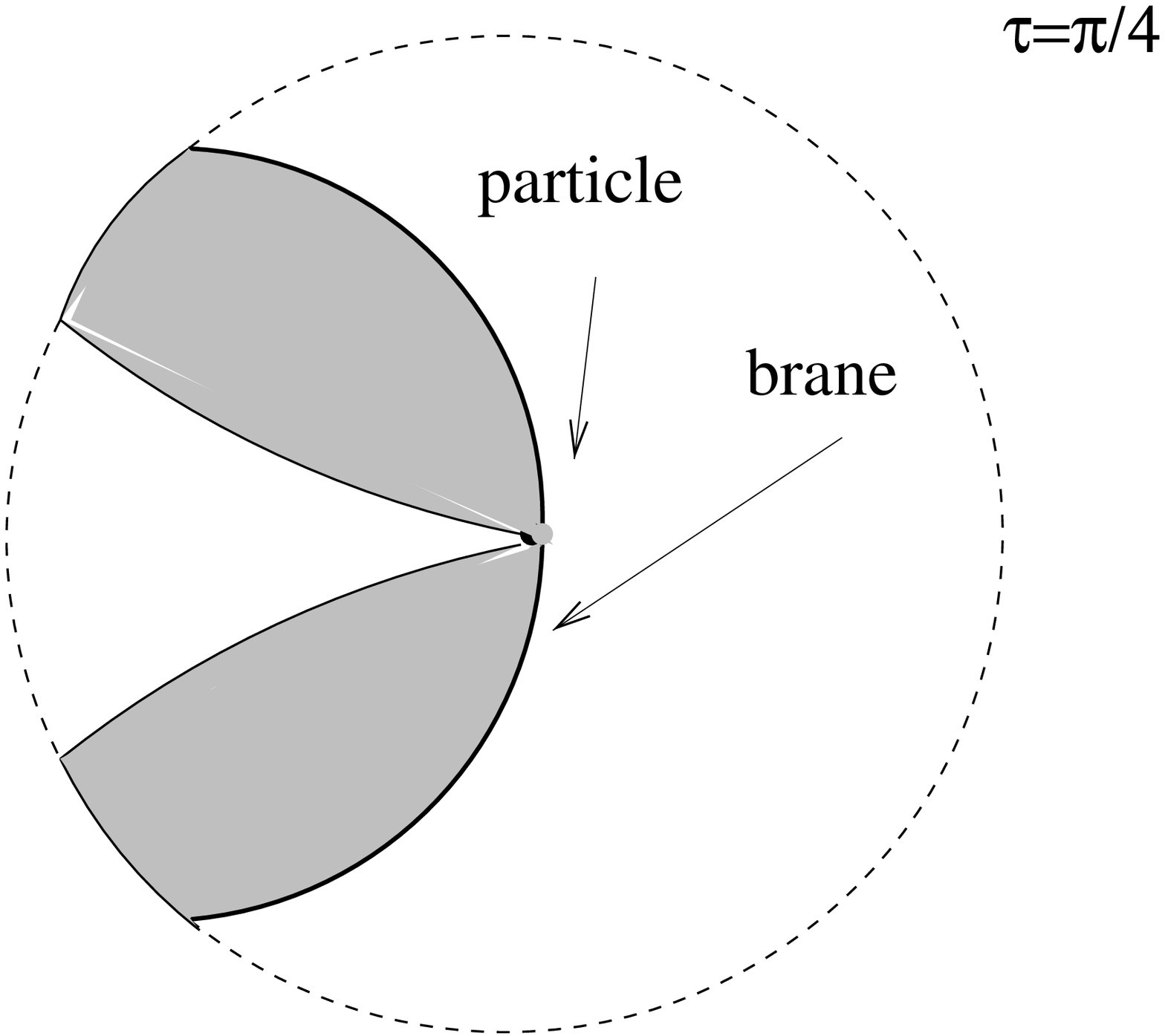,
         width=100pt,
                          angle=0}$~~~~~~~~~~~~~$
\epsfig{file=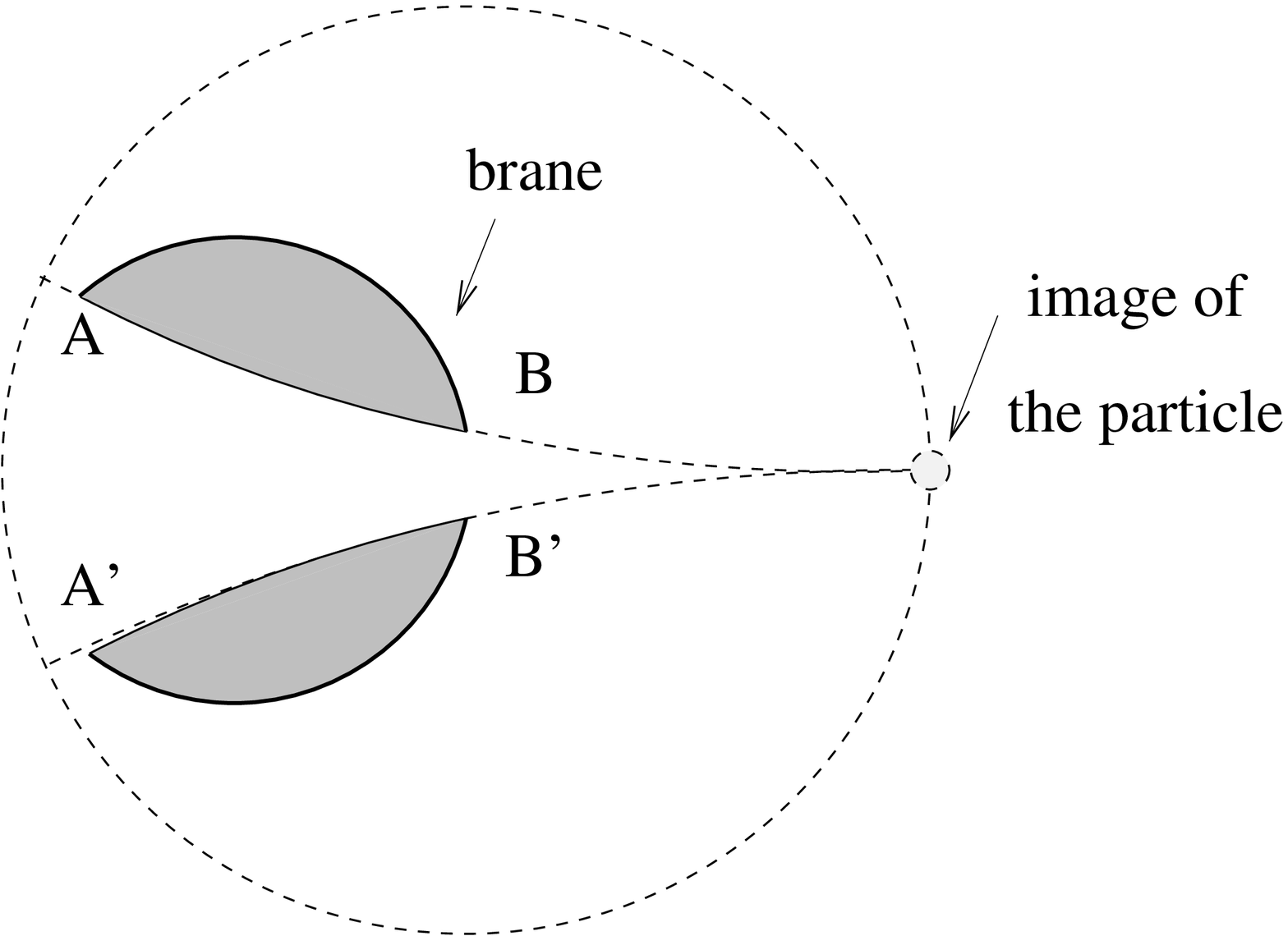,
   width=115pt,
  angle=0
 }
\end{center}
\label{Fig6}
\caption{Brane positions on the Poincare disk with wedge cuts ($\epsilon
=1/12 \pi)$ in  times
$-\pi/2\leq
\tau \leq \pi/2$}
\end{figure}
\begin{figure}[h]
\begin{center}
\epsfig{file=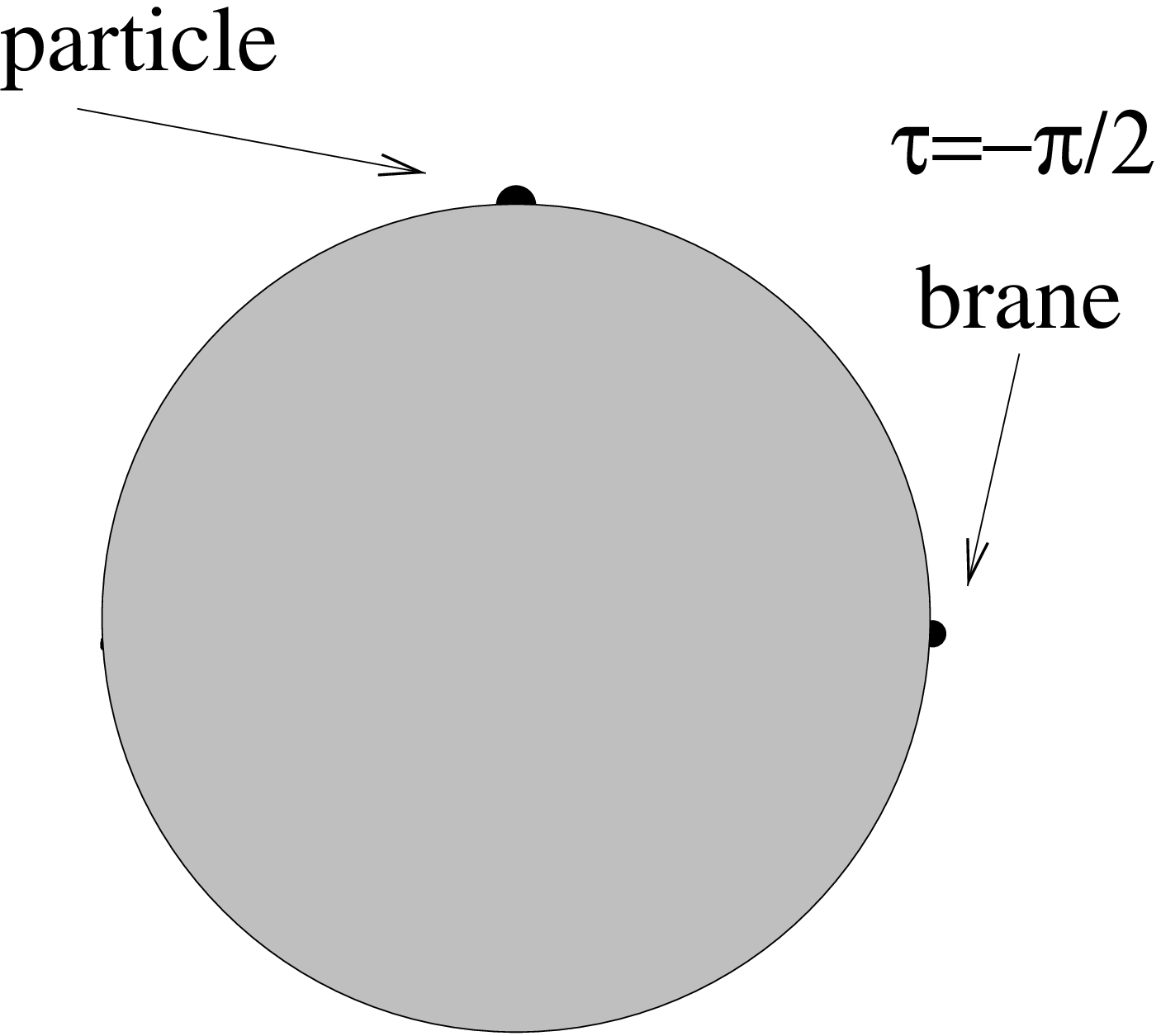,
           width=120pt,
           angle=0}$~~~~~~~~~~~~~$
\epsfig{file=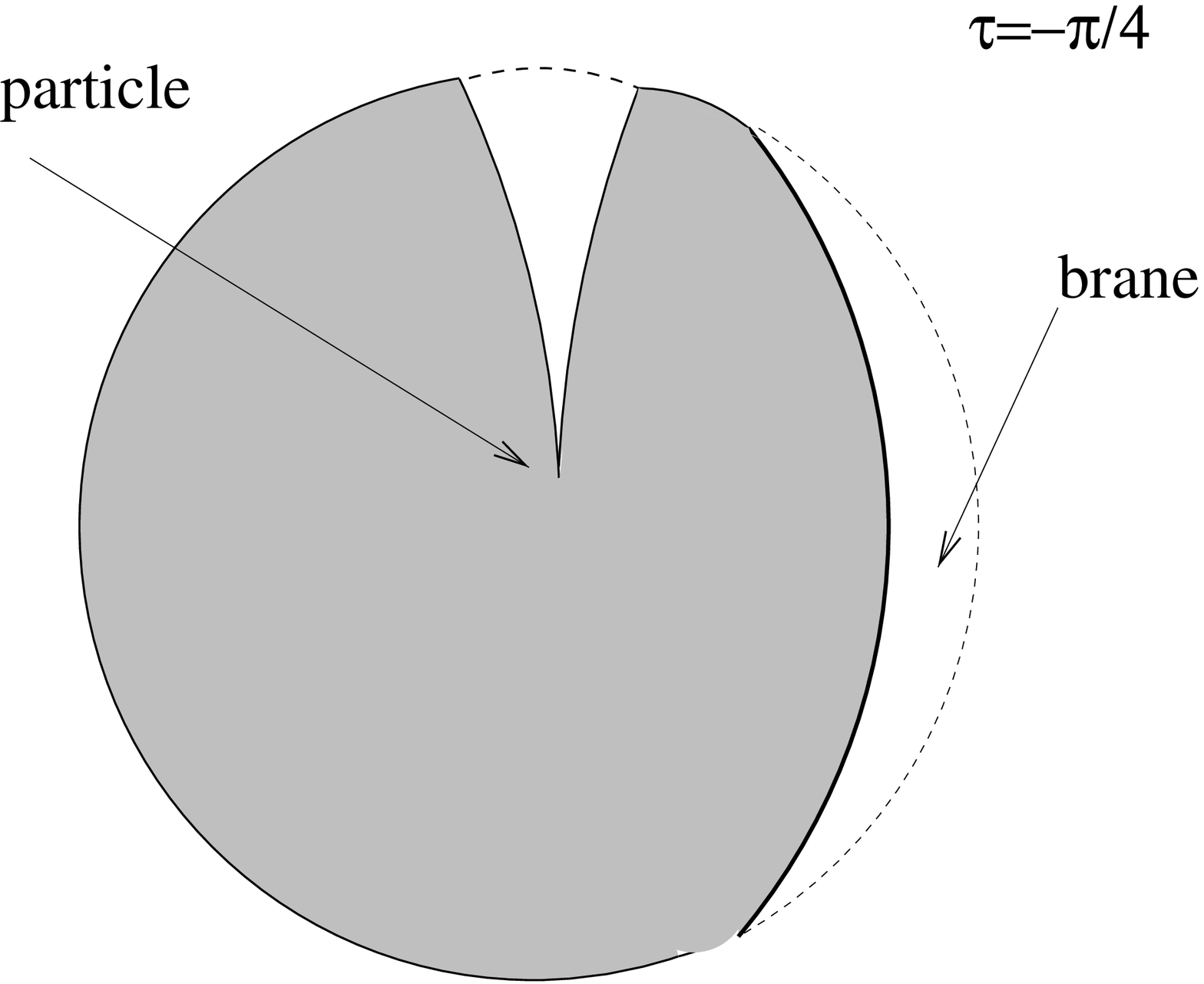,
         width=110pt,
                      angle=0}$~~~~~~~~~~~~~$
\epsfig{file=
         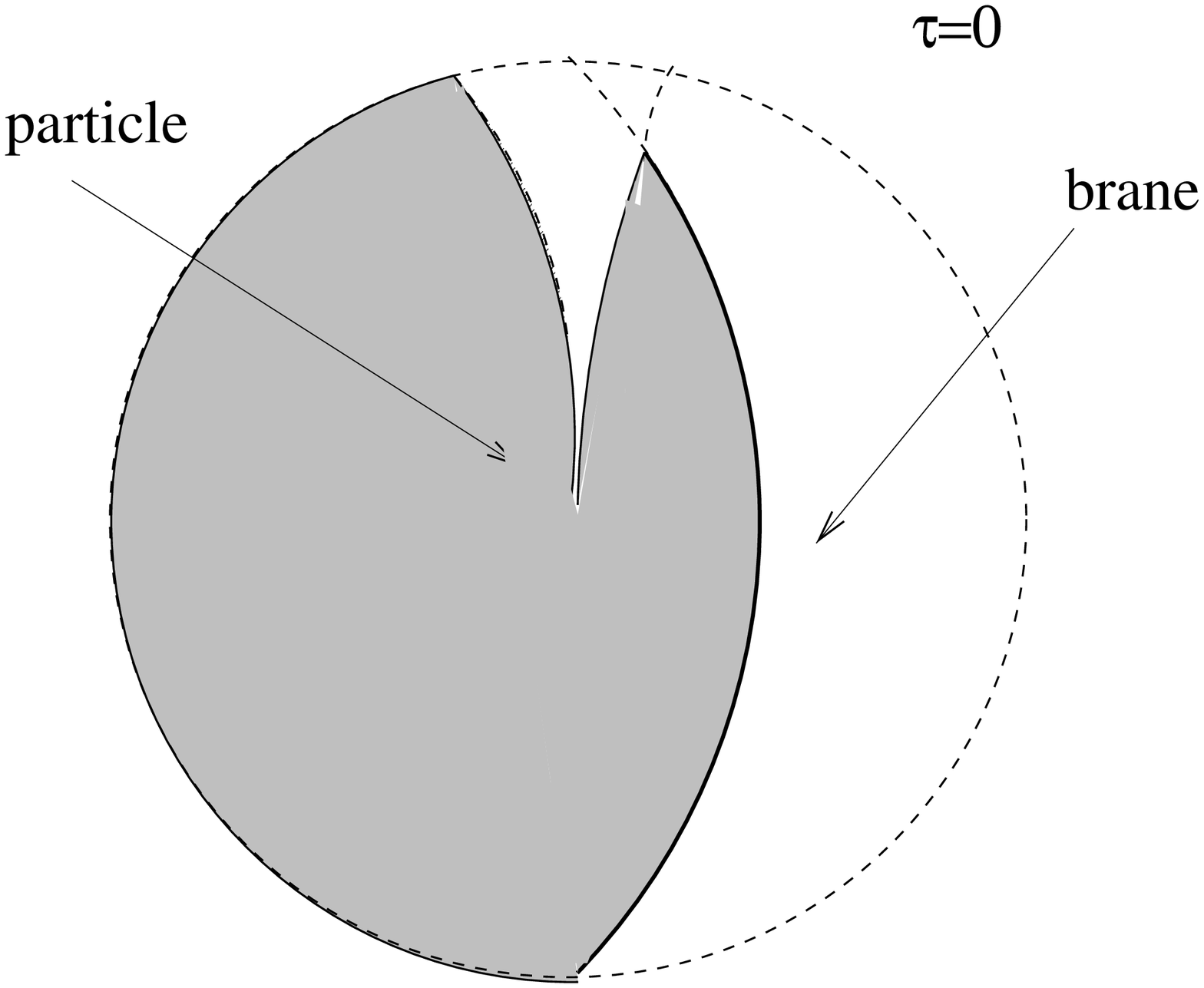,
         width=110pt,
                                 angle=0}
\end{center}
$$~$$
\begin{center}
\epsfig{file=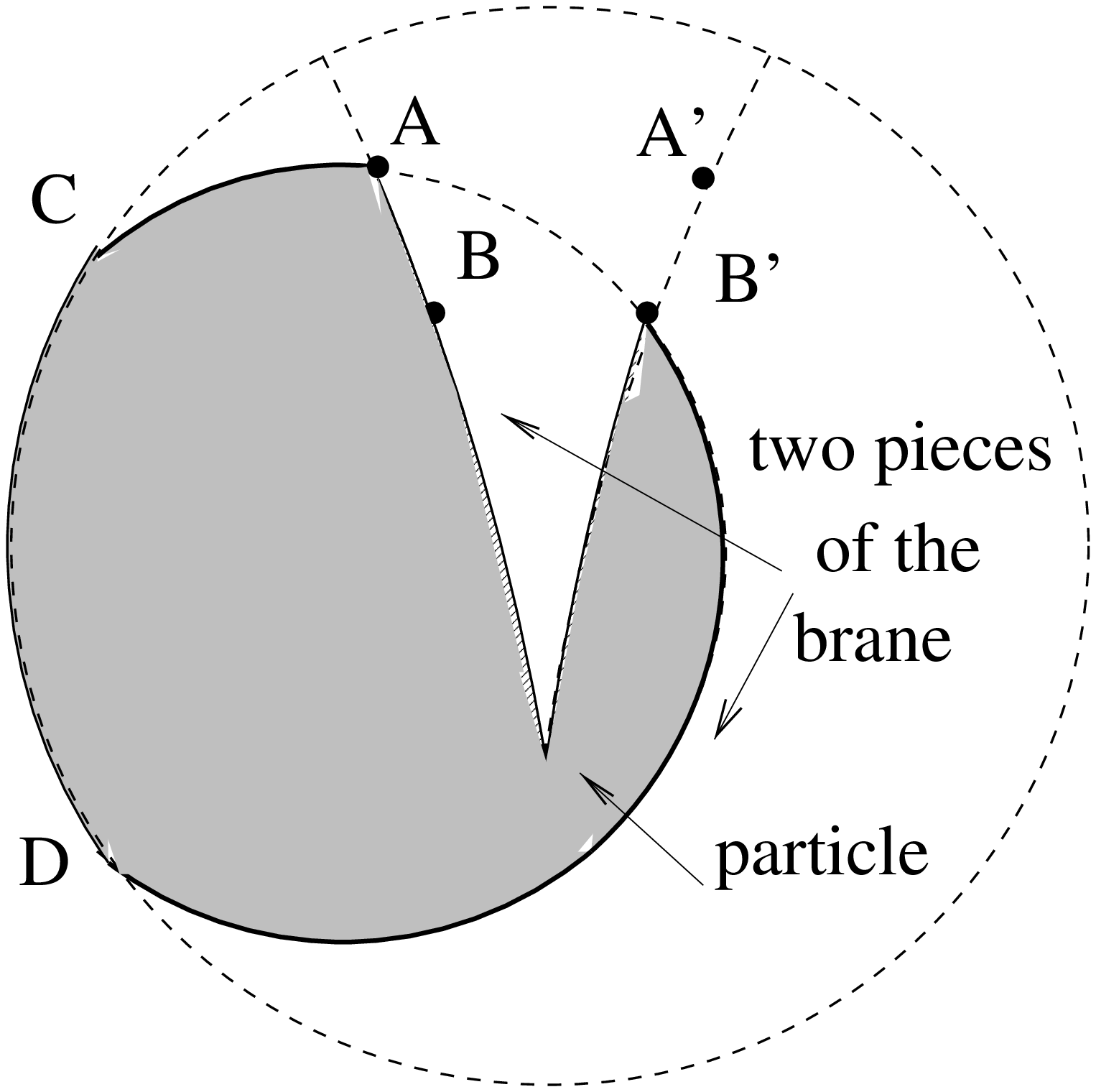,
         width=90pt,
                          angle=0}$~~~~~~~~~~~~~$
\epsfig{file=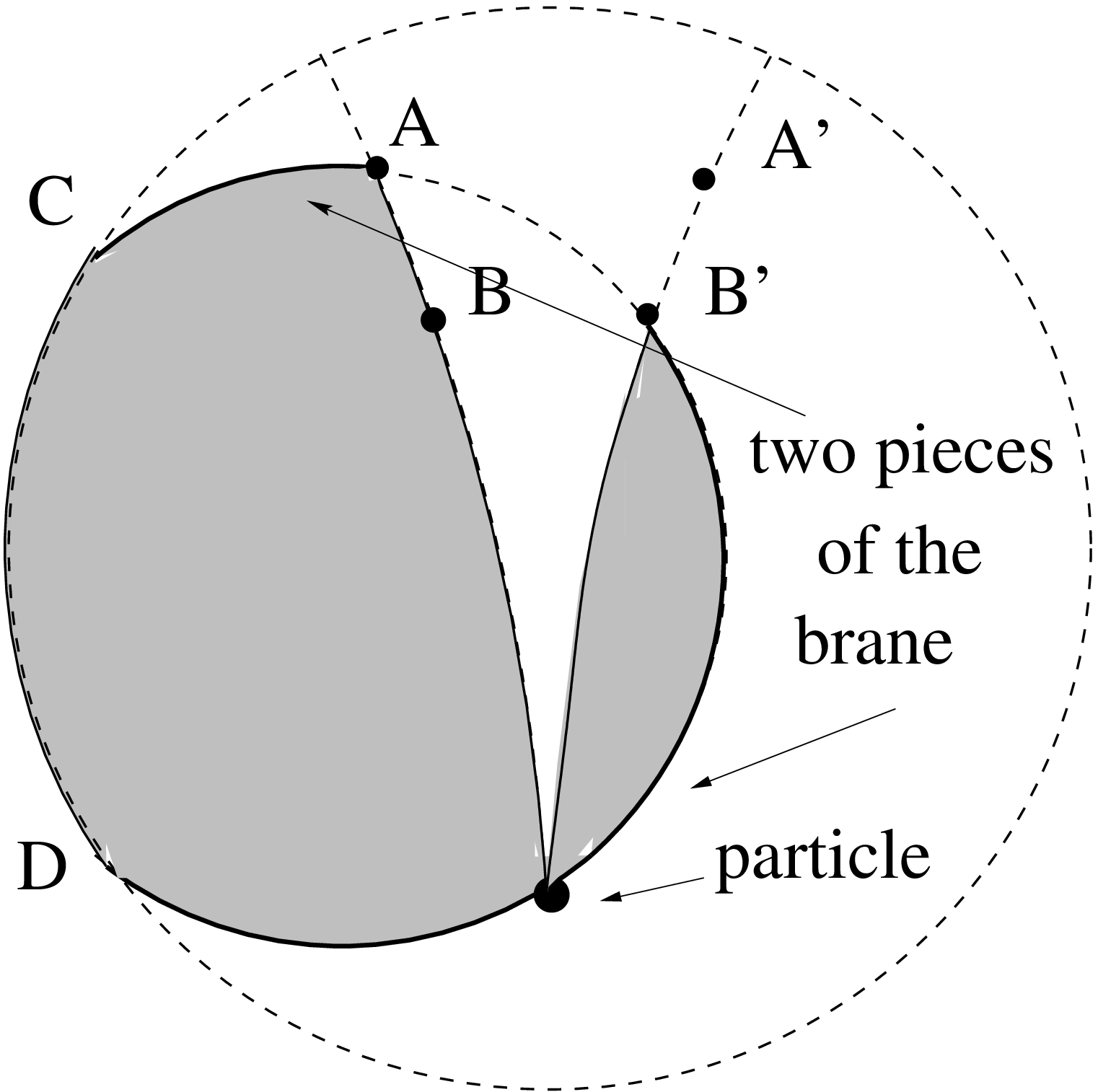,
 width=90pt,
        angle=0}$~~~~~~~~~~~~~$
\epsfig{file=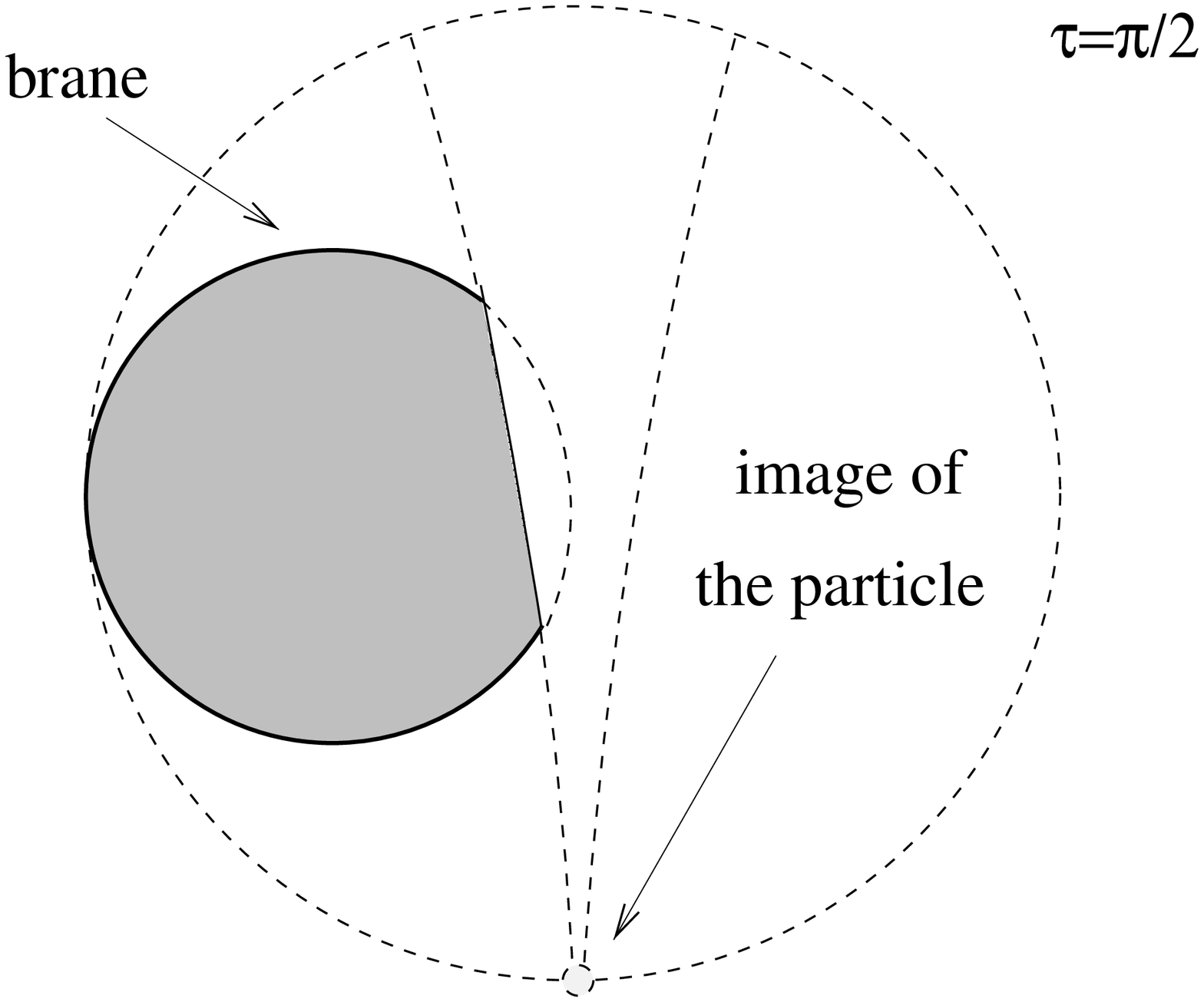,
   width=110pt,
  angle=0
 }
\end{center}
\label{Fig7}
\caption{Brane positions on the Poincare disk with wedge cuts ($\epsilon
=1/12 \pi)$ in  times
$-\pi/2\leq
\tau \leq \pi/2$}
\end{figure}
\end{document}